%% file: main.tex
\documentclass[lettersize,journal]{IEEEtran}
\usepackage{amsmath,amsfonts}

\usepackage{algorithmic}
\usepackage{algorithm}
\usepackage{array}
\usepackage[caption=false,font=normalsize,labelfont=sf,textfont=sf]{subfig}
\usepackage{textcomp}
\usepackage{stfloats}
\usepackage{url}
\usepackage{verbatim}
\usepackage{graphicx}
\usepackage{cite}
\usepackage{booktabs}
\usepackage[colorlinks,linkcolor=red,citecolor=green,urlcolor=blue]{hyperref}
\usepackage{multirow}
\usepackage{makecell}
\begin{document}

\title{Multimodal Classification and Out-of-distribution Detection for Multimodal Intent Understanding}

\author{
        Hanlei Zhang*,
        Qianrui Zhou*,
        Hua Xu,~\IEEEmembership{Member,~IEEE,}
        Jianhua Su,
        Roberto Evans,
        Kai Gao
\thanks{
This work was supported in part by the National Natural Science Foundation of China under Grant 62173195; in part by the National Science Technology Major Project towards the New Generation of Broadband Wireless Mobile Communication Networks of Jiangxi Province (03, 5G Major Project of Jiangxi Province) under Grant 20232ABC03402; in part by the High-level Scientific Technological Innovation Talents “Double Hundred Plan” of Nanchang City in 2022 under Grant Hongke Zi (2022) 321-16; and in part by the Natural Science Foundation of Hebei Province, China, under Grant F2022208006. Hua Xu is the corresponding author.\\
Hanlei Zhang, Qianrui Zhou, Hua Xu, and Roberto Evans are with the State Key Laboratory of Intelligent Technology and Systems, Department of Computer Science and Technology, Tsinghua University, Beijing 100084, China (e-mail: zhang-hl20@mails.tsinghua.edu.cn; zgr22@mails.tsinghua.edu.cn; xuhua@tsinghua.edu.cn; linyf21@mails.tsinghua.edu.cn).\\
Jianhua Su is with the State Key Laboratory of Intelligent Technology and Systems, Department of Computer Science and Technology, Tsinghua University, Beijing 100084, China; with Samton (Jiangxi) Technology Development Company Ltd., Nanchang, China; and with the School of Information Science and Engineering, Hebei University of Science and Technology, Shijiazhuang 050018, China (e-mail: sjh19832107090@gmail.com).\\
Kai Gao is with the School of Information Science and Engineering, Hebei University of Science and Technology, Shijiazhuang, Hebei 050018, China (e-mail: gaokai@hebust.edu.cn).
} 
\thanks{*: These authors contributed equally to this work.}
}
        



\maketitle

\input{Abstract}

\begin{IEEEkeywords}
Multimodal intent understanding, intent classification, out-of-distribution detection, multimodal fusion. 
\end{IEEEkeywords}

\section{Introduction}

Multimodal intent understanding is a rapidly growing field in multimodal language analysis~\cite{zhang2022mintrec}. It aims to leverage information from various modalities in real-world scenarios (i.e., text, video, and audio) to enable machines to comprehend complex semantics (e.g., intents) from human conversations. Recognizing user intents can significantly enhance service quality and can be applied in numerous substantial applications, such as virtual humans~\cite{Wang2023Multimodal}, chatbots~\cite{10.1145/3488560.3502189}, and other interaction systems~\cite{10069301, Zhang2020Towards, Mi2019Object}.

\begin{figure}
\centering
\includegraphics[scale=.35]{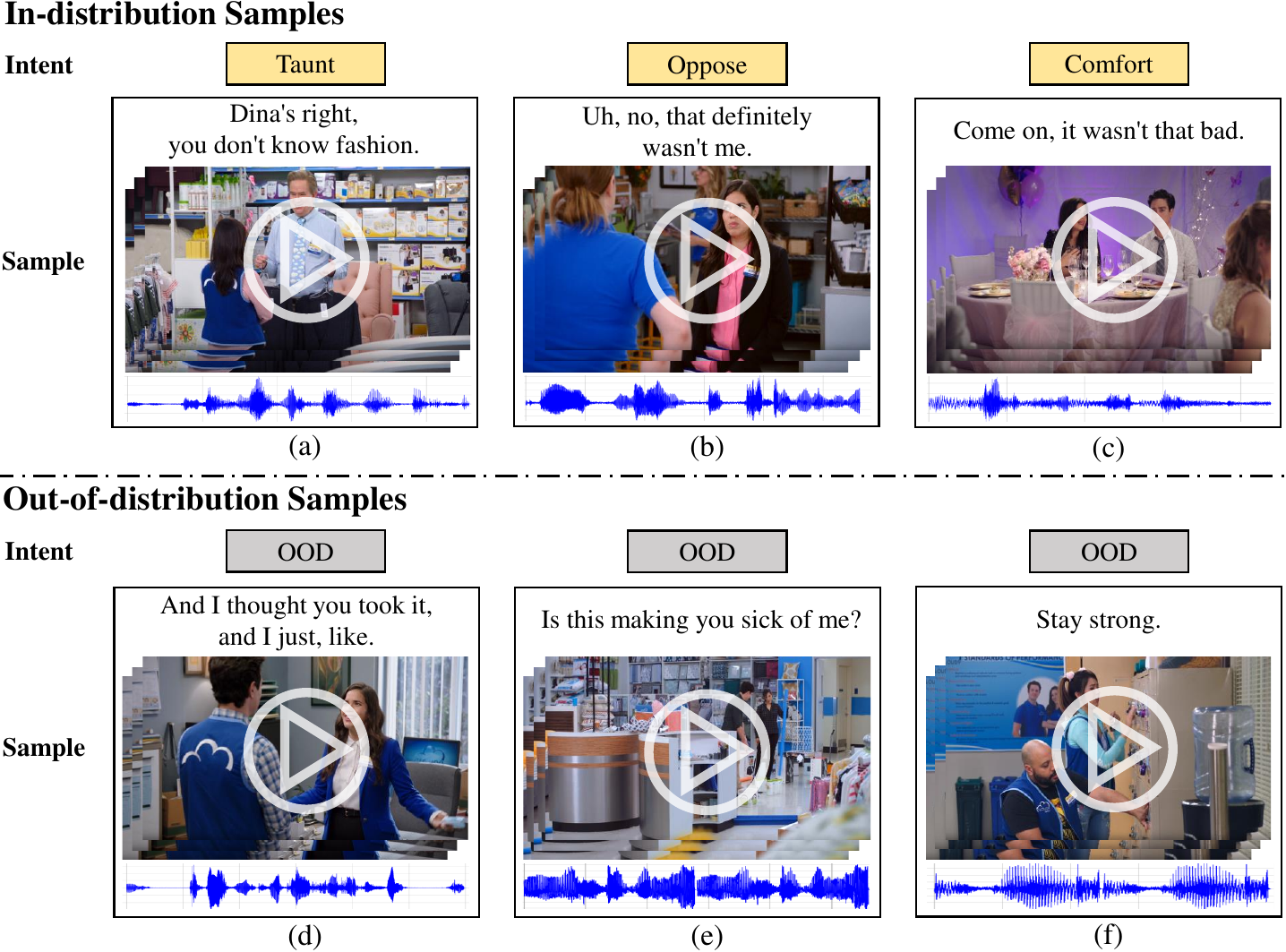}
\caption{Examples of in-distribution and out-of-distribution samples for multimodal intent understanding.}
\label{Introduction}
\vspace{-0.4cm}
\end{figure}

However, research in multimodal intent understanding is still in its early stages, and related studies have just begun in recent years. One challenge is how to effectively perform multimodal fusion and learn representations conducive to intent classification. For example, Zhang et al.~\cite{zhang2022mintrec} introduced the first multimodal intent dataset with annotations from 20 intent categories and establishes baselines using multimodal fusion methods~\cite{tsai2019multimodal,hazarika2020misa,rahman2020integrating} adapted from the multimodal sentiment analysis area, demonstrating the effectiveness of non-verbal modalities in this task. Huang et al.~\cite{huang2024sdif} introduced an interaction framework that aligns video and audio with the text modality before fusing all multimodal features using a transformer. Zhou et al.~\cite{zhou2024token} proposed a prompt-based module to construct multimodal augmented pairs for token-level contrastive learning. However, these methods have limitations in learning discriminative representations to distinguish fine-grained and semantically complex intent classes.

Another challenge is that existing methods are limited to closed-world classification on in-distribution (ID) intent data and struggle with out-of-distribution (OOD) data from unseen open classes~\cite{Zhang_Xu_Lin_2021,10097558} or irrelevant out-of-scope utterances~\cite{larson-etal-2019-evaluation}, commonly encountered during conversational interactions in real-world scenarios. Figure~\ref{Introduction} illustrates examples of ID and OOD data for multimodal intent understanding. Both ID and OOD data contain text, video, and audio modalities, and we combine these modalities to analyze the intents. Specifically, ID data come from predefined known intent classes, such as \textit{Taunt}, \textit{Oppose}, and \textit{Comfort}, listed in figures (a), (b), and (c), respectively. Though the text modality is predominant in this task~\cite{zhou2024token}, non-verbal modalities often play enhancing or necessary roles. For instance, in example (a), using only the text modality might infer the speaker's intention as \textit{Criticize}, but analyzing the speaker's expressions and tones from non-verbal modalities corrects it to \textit{Taunt}. OOD data consist of multimodal utterances that do not belong to existing known intents, including potential new intents, such as~\textit{Doubt} and~\textit{Encourage} in examples (e) and (f), and statements of one's opinion as in example (d). These OOD samples are commonly encountered in daily interactions, but most existing OOD detection methods focus on a single modality, such as vision~\cite{9156473, liu2020energy}, text~\cite{zhou-etal-2022-knn, zhan-etal-2021-scope}, or leveraging one modality as prior knowledge to help the target modality for OOD detection~\cite{esmaeilpour2022zero,ming2022delving}. These methods fail to effectively leverage multiple modalities to generate cooperative representations for OOD detection. When adapting existing multimodal fusion methods to OOD detection, our experiments show that they tend to overfit on ID classes and lack robustness and generalization when encountering unseen OOD samples.

To address these two challenges, we propose MIntOOD, a novel method for multimodal intent classification and OOD detection. The motivation is that both tasks can be improved through effective modeling of multimodal representations and learning from discriminative information among both ID and pseudo-OOD data at multiple granularity levels. For modeling multimodal representations, we first extract feature embeddings for each modality with advanced backbones in their respective fields and then construct pseudo-OOD features of each modality through convex combinations of a certain amount of ID features from at least two different classes following the Dirichlet distribution. Next, we build encoders for each modality and design a weighted feature fusion network for effective multimodal fusion. In particular, the importance of each modality is considered by dynamically learning corresponding weights adaptive to multimodal contexts via neural networks. We perform a weighted summation on the encoded representations and normalized weights of each modality.

To learn discriminative and robust multimodal representations for both ID classification and OOD detection, we optimize the learning objectives from three perspectives. First, we use the ID and pseudo-OOD features to learn coarse-grained distinguishing characteristics by differentiating whether a sample is ID or OOD. Second, we employ a cosine classifier to enhance discrimination among specific ID classes by eliminating the effect of vector magnitude and focus on angular deviations to improve class separability. Finally, we focus on a more fine-grained perspective within instance-level ID and OOD samples. We apply contrastive learning to pull ID samples with the same class together and push away samples from different ID classes to learn ID instance interactions. When involving OOD, we force each OOD sample to separate from any other samples to enhance the model's discrimination capability on various constructed OOD samples, thereby improving its generalization ability and robustness when encountering unseen OOD data.

Our contributions are summarized as follows: (1) We introduce MIntOOD, a novel method for multimodal intent classification and OOD detection. To the best of our knowledge, this is the first multimodal method that shows strong generalization ability on unseen OOD data while ensuring ID classification performance. (2) We achieve a simple and effective multimodal fusion network by fusing the encoded representations of each modality with corresponding importance degrees, through dynamically learning weight scores adaptive to multimodal contexts. (3) We design optimization objectives for learning robust representations from three different granularities, fully capturing both coarse-grained and fine-grained distinction information within ID and constructed OOD data. (4) We establish baselines for three multimodal intent datasets and create an OOD benchmark. Extensive experiments show that our method achieves new state-of-the-art performance on ID classification and significant improvements on OOD detection of 3\%$\sim$10\% on AUROC over the best-performing baselines.

\section{Related Works}
This section reviews the literature on intent understanding, multimodal fusion, and out-of-distribution detection. 

\subsection{Intent Understanding}
Intent understanding is a significant research area originating from natural language understanding (NLU) and aims to analyze the semantics underlying dialogue utterances, a field that has been gaining popularity~\cite{ijcai2021p622}. Researchers have proposed various benchmark datasets~\cite{coucke2018snips,larson-etal-2019-evaluation,casanueva-etal-2020-efficient} to advance the field. For these benchmark datasets, traditional supervised learning methods have achieved superior performance~\cite{chen2019bert,he2022unified}. However, OOD samples frequently appear during dialogue interactions, prompting research into open-world intent analysis, including open intent detection~\cite{Zhang_Xu_Lin_2021,10097558,zhou-etal-2022-knn} and new intent discovery~\cite{lin2020discovering,zhang2021discovering,zhou2023probabilistic,zhang2023clustering}. The former trains on data from ID classes to recognize these known classes while also detecting unseen open classes during testing. The latter uses unsupervised or semi-supervised data to identify potential intent groups. However, these works focus solely on the text modality and fail to leverage non-verbal modalities, which are crucial in dialogue interactions.

Recently, there has been increasing interest in exploring intent understanding in multimodal scenarios. For instance, Kruk et al.~\cite{kruk2019integrating} introduced the MDID dataset, which combines image-text pairs from Instagram posts to analyze authors’ intents, while Zhang et al.~\cite{zhang2021multimodal} studied market intent understanding using text and image modalities from social news. However, these tasks differ from the multimodal intent analysis found in conversational settings. Singh et al.~\cite{singh2022emoint} proposed a multimodal contextual transformer network that encodes utterances with contextual information from each modality and captures modality-specific and modality-invariant properties. The MIntRec dataset~\cite{zhang2022mintrec} makes a pioneering contribution in this area, including 2,224 multimodal utterances containing text, video, and audio modalities annotated from 20 intent categories. Moreover, it provides baselines with powerful multimodal fusion methods adapted from multimodal sentiment analysis. Based on this dataset, Yu et al.~\cite{yu2023speech} proposed a bi-modality pre-trained model that aligns speech and text by predicting temporal positions and selecting cross-modal responses. Moreover, Huang et al.~\cite{huang2024sdif} designed a shallow-to-deep interaction module that aligns non-verbal modalities with text through a shallow interaction layer. It then fuses all modalities using a transformer-based deep interaction module. Additionally, it augments the training data from large language models to further improve performance. Dialogue act is a type of communicative intent~\cite{firdaus2021deep} that is more coarse-grained than the specific intent classes defined in~\cite{zhang2022mintrec}. However, previous dialogue act datasets contain only the text modality~\cite{225858,li2017dailydialog}. Saha et al.~\cite{saha-etal-2020-towards} introduced the first multimodal dialogue act dataset, derived from MELD~\cite{poria2019meld} and IEMOCAP~\cite{busso2008iemocap}, adding dialogue act annotations and proposing a triplet attention subnetwork to capture intra- and inter-modality characteristics. In this work, we use MIntRec, MELD-DA, and IEMOCAP-DA as benchmark datasets for multimodal intent recognition. The current state-of-the-art method, TCL-MAP~\cite{zhou2024token}, aligns non-verbal modalities with text, generates modality-aware prompts using a cross-modal transformer, and employs contrastive learning on masking tokens with the help of the generated prompts. While it achieves strong performance on the ID classes of MIntRec and MELD-DA, it demonstrates poor generalization on unseen OOD data in our experiments.

\subsection{Multimodal Fusion} 
With the emergence of multimodal language datasets in recent years~\cite{yu2020ch,zadeh2016mosi,zadeh2018multimodal,zhang2022mintrec,zhang2024mintrec}, there has been significant interest in exploring multimodal fusion techniques, aiming to combine features from multiple modalities to capture their interactions. Early methods used tensor operations for multimodal fusion, relying on the rich semantics in high-dimensional representations~\cite{zadeh-etal-2017-tensor, liu-etal-2018-efficient-low}. However, these methods struggle to balance computational cost with model performance. To address this issue, Zadeh et al.~\cite{zadeh2018memory} provided a solution by learning view-specific interactions and employing an attention mechanism with a multi-perspective gated memory.

In recent years, researchers have turned to transformer-based methods for multimodal fusion, owing to their excellent performance and efficiency in natural language processing and computer vision. For example, Tsai et al.~\cite{tsai2019multimodal} used directional pairwise cross-modal attention for fusing multimodal sequences to adapt information streams from one modality to another. Rahman et al.~\cite{rahman2020integrating} attached a multimodal adaptation gate to transformers, allowing it to receive information from different modalities during fine-tuning. Hazarika et al.~\cite{hazarika2020misa} separated each modality into modality-invariant and modality-specific subspaces for fine-grained fusion. Han et al.~\cite{han2021improving} designed a hierarchical mutual information maximization method that successively maximizes in inter-modality and fusion levels. Hu et al.~\cite{hu2022unimse} proposed a general framework that unifies both multimodal emotion recognition and sentiment analysis tasks, formulating these tasks into a universal label format and incorporating a multimodal fusion layer into the pre-trained model T5~\cite{raffel2020exploring}, learning inter-modality interactions with contrastive learning. However, the required label formats are not suitable for our task.

\subsection{Out-of-distribution Detection}  
Research on out-of-distribution (OOD) detection can be broadly categorized into two main aspects. On one hand, researchers have focused on learning robust representations for achieving better OOD detection performance. For example, Zhou et al.~\cite{zhou-etal-2022-knn} attempted to learn more robust representations of IND data. Zheng et al.~\cite{zheng2020out} utilized unlabeled data to generate pseudo-OOD data. Lee et al.~\cite{lee-etal-2023-improving} leveraged latent categorical information by training a transformer-based sentence encoder for pseudo-labeling and then performing pseudo-label learning. Shen et al.~\cite{shen2021enhancing} designed a domain-regularized module to alleviate overconfidence in classifiers. Zhang et al.~\cite{zhan-etal-2022-closer} proposed a two-stage method that first learns to generate synthetic ID samples for data enrichment and then adopts a training approach with \textit{K}+1 classes for discriminative training in OOD detection. However, these methods are constrained to the text modality and fail to leverage non-verbal modalities.

On the other hand, many works have designed scoring functions to calibrate model outputs for OOD detection. For example, Liu et al.~\cite{liu2020energy} applied the energy score by calculating the logsumexp on logits. Lee et al.~\cite{lee2018simple} used the minimum Mahalanobis distance between features and class centroids as a confidence score. Ndiour et al.~\cite{ndiour2020out} focused on the residual norm between a feature and its low-dimensional embedded pre-image. Wang et al.~\cite{wang2022vim} combined information from both features and categories, using the residual norm in~\cite{ndiour2020out} as a virtual logit component to calculate its softmax probability. In this work, we directly apply these scoring functions after the model outputs and compare the performance in Section~\ref{OOD_comparison}.

\begin{figure*}
	\centering
	\includegraphics[scale=0.9]{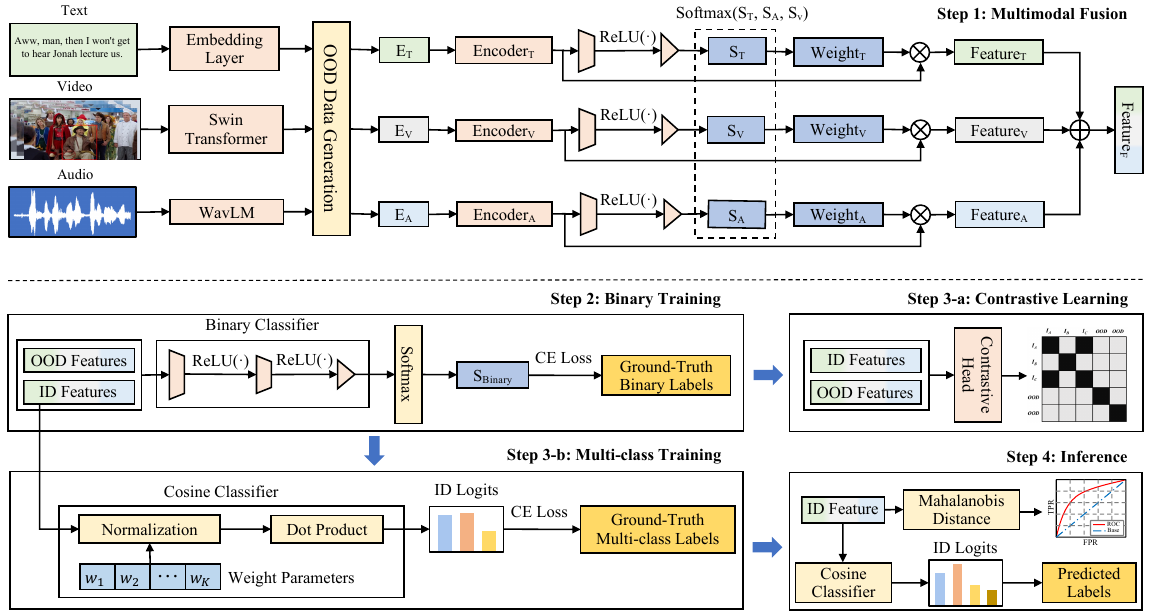}
	\caption{Overall architecture of MIntOOD. It begins by generating pseudo-OOD samples through convex combinations of features extracted from ID samples. A weighted feature fusion network is then designed to dynamically learn an importance score for each modality, serving as weights during feature fusion. To learn robust representations for both ID classification and OOD detection, MIntOOD focuses on three granularities: (a) coarse-grained binary information, learned via a binary classifier trained to distinguish ID or OOD classes, (b) fine-grained information for distinguishing ID classes, captured by using a cosine classifier to analyze the angular information, and (c) fine-grained information for distinguishing instance-level differences, achieved through contrastive learning that captures similarity relations among ID and OOD utterances.}
	\label{MIntOOD}
\end{figure*}

\section{The Proposed Approach: MIntOOD}
This section formulates the problem of multimodal intent classification and OOD detection, and introduces a novel method, MIntOOD, as illustrated in Figure~\ref{MIntOOD}. 

\subsection{Problem Formulation}
Consider an intent dataset $\mathcal{D} = \{\mathcal{D}_{\textrm{ID}}, \mathcal{D}_{\textrm{OOD}}\}$, where $\mathcal{D}_{\textrm{ID}}$ and $\mathcal{D}_{\textrm{OOD}}$ denote the ID and OOD data, respectively. The ID data appears in the training, validation, and testing sets, while the OOD data only appears in the testing set.

\textbf{ID Classification.} Let $\mathcal{D}_{\text{ID}} = \{(x_i, y_i) \mid y_{i} \in \mathcal{I} \}_{i=1}^{N_{\textrm{ID}}}$ denote the ID dataset consisting of $N_{\textrm{ID}}$ instances. Here, $x_i = (x_i^{\textrm{T}}, x_i^{\textrm{V}}, x_i^{\textrm{A}})$ represents the $i^{\textrm{th}}$ utterance with text, video, and audio modalities, $\mathcal{I} = \{\mathcal{I}\}_{i=1}^{K}$ is the set of intent labels, and $K$ is the number of intent classes. The goal is to accurately predict the label $y_{i}$ for each utterance $x_{i}$.

\textbf{OOD Detection.} Let $\mathcal{D}_{\text{OOD}} = \{(x_j, y_j) \mid y_j = y_{\text{OOD}},\, y_j \notin \mathcal{I}\}_{j=1}^{N_{\textrm{OOD}}}$ denote the OOD dataset containing $N_{\textrm{OOD}}$ utterances that do not belong to any known class. $y_{\textrm{OOD}}$ is a unified special token marking the labels of all OOD data. The goal is to distinguish OOD utterances from ID utterances during testing.

\subsection{Feature Extraction and OOD Data Generation}
\textbf{Feature Extraction.} Following~\cite{zadeh2018multimodal,zhang2022mintrec}, we first extract feature embeddings for text, video, and audio modalities to facilitate multimodal fusion. For each text utterance $x^{\mathrm{T}}$, we obtain the embeddings 
\[
\mathbf{E}^{\mathrm{T}}=\{\mathbf{E}^{\mathrm{T}}_{[\text{CLS}]}, \mathbf{E}^{\mathrm{T}}_{1}, \ldots, \mathbf{E}^{\mathrm{T}}_{[\text{SEP}]}\}\in\mathbb{R}^{L_{\mathrm{T}} \times D_{\mathrm{T}}}
\]
by summing the corresponding token, segment, and position embeddings from a pre-trained BERT language model~\cite{chen2019bert}. For each video segment $x^{\mathrm{V}}$, we acquire the image sequence embeddings $\mathbf{E}^{\mathrm{V}}\in\mathbb{R}^{L_{\mathrm{V}} \times D_{\mathrm{V}}}$ using the Swin Transformer backbone~\cite{liu2021swin}, as validated in~\cite{zhou2024token}. For audio, we first process waveform data with the librosa toolkit~\cite{mcfee2015librosa}, following~\cite{zhang2022mintrec}, and then use the WavLM model~\cite{chen2022wavlm} to extract audio sequence embeddings $\mathbf{E}^{\mathrm{A}}\in\mathbb{R}^{L_{\mathrm{A}} \times D_{\mathrm{A}}}$. Here, $L_{\mathrm{T}}$, $L_{\mathrm{V}}$, and $L_{\mathrm{A}}$ denote the sequence lengths for text, video, and audio, respectively, while $D_{\mathrm{T}}$, $D_{\mathrm{A}}$, and $D_{\mathrm{V}}$ indicate the dimensionalities of these modalities. All multimodal fusion methods use the same tri-modal embeddings as inputs to ensure a fair comparison.

\textbf{OOD Data Generation.} Obtaining high-quality OOD data in real scenarios is often prohibitively expensive~\cite{ouyang2021energy}. To reduce this cost, we generate pseudo-OOD data by leveraging the ID embeddings of each modality (i.e., $\mathbf{E}^{\mathrm{T}}_{\textrm{ID}}$, $\mathbf{E}^{\mathrm{A}}_{\textrm{ID}}$, and $\mathbf{E}^{\mathrm{V}}_{\textrm{ID}}$), thereby helping the model learn to differentiate between ID and OOD. Specifically, we randomly select $k$ embedded examples in each training batch, ensuring they are drawn from at least two distinct classes. We then create convex combinations of these selected samples to generate the $i^{\textrm{th}}$ OOD embedding:
\begin{align}
\mathbf{E}^{\mathrm{M}}_{\textrm{OOD},i} &= \sum_{j=1}^{k}\lambda_{j} \cdot \mathbf{E}^{\mathrm{M}}_{\textrm{ID},j},  \label{OOD_embedding_formula}\\
\text{s.t. } \lambda &\sim \textrm{Dir}(\alpha), \quad |\mathcal{C}(\{y_j\}_{j=1}^k)| \geq 2,
\end{align}
where $\mathrm{M} \in \{\mathrm{T}, \mathrm{V}, \mathrm{A}\}$ indicates the modality. We sample the coefficients $\{\lambda_j\}_{j=1}^{k}$ from a Dirichlet distribution $\textrm{Dir}(\alpha)$ such that $\sum_{j=1}^{k} \lambda_{j} = 1$ and each $\lambda_{j} \in [0,1]$. Here, $\alpha$ is a hyper-parameter that controls the diversity of the pseudo-OOD data, and $\mathcal{C}(\cdot)$ denotes the unique classes among the selected samples’ labels.

Compared with existing manifold mixup strategies~\cite{9693239,zhan-etal-2021-scope}, which generally rely on binary sample mixing, our method introduces more diversity by sampling from a Dirichlet distribution among multiple samples. This expands the range of OOD embeddings, improving the model’s generalization across various scenarios. Furthermore, it mitigates the risk of overfitting to ID data and promotes more robust learning, ultimately boosting performance on unseen OOD data.

\subsection{Multimodal Fusion}
\textbf{Multimodal Encoders.} To capture high-level semantic relationships and temporal information in text, video, and audio, we employ a multi-head attention mechanism within transformer-based encoders. For the text modality, we use a pre-trained BERT~\cite{devlin-etal-2019-bert}:
\begin{align}
    \mathbf{x}^{\mathrm{T}} = \textrm{BERT}(\mathbf{E}^{\mathrm{T}})_{[\textrm{CLS}]},
\end{align}
where $\mathbf{x}^{\mathrm{T}} \in \mathbb{R}^{D_{\mathrm{T}}}$. The $[\textrm{CLS}]$ token from the final hidden layer serves as the sentence-level representation~\cite{zhang2022mintrec}. We then apply vanilla transformers~\cite{vaswani2017attention} to obtain video and audio features:
\begin{align}
    \mathbf{x}^{\mathrm{A}} &= \mathbf{W}_{e}^{\mathrm{A}}\bigl(\operatorname{mean-pooling}(\textrm{Transformer}_{\mathrm{A}}(\mathbf{E}^{\mathrm{A}}))\bigr),\\
    \mathbf{x}^{\mathrm{V}} &= \mathbf{W}_{e}^{\mathrm{V}}\bigl(\operatorname{mean-pooling}(\textrm{Transformer}_{\mathrm{V}}(\mathbf{E}^{\mathrm{V}}))\bigr),
\end{align}
where $\mathbf{W}_{e}^{\mathrm{A}} \in \mathbb{R}^{D_{\mathrm{A}} \times D_{\mathrm{T}}}$ and $\mathbf{W}_{e}^{\mathrm{V}} \in \mathbb{R}^{D_{\mathrm{V}} \times D_{\mathrm{T}}}$. As the text modality has explicit token-level embeddings, we use its $[\textrm{CLS}]$ vector directly. In contrast, video and audio representations are derived by mean-pooling the respective transformer outputs, then mapping them to the text-dimensional space.

\textbf{Weighted Feature Fusion Network.} After encoding each modality into semantically rich representations $\mathbf{x}^{\{\mathrm{T, A, V}\}}$, we explore multimodal fusion. Many traditional methods simply concatenate or add these embeddings~\cite{tsai2019multimodal,hazarika2020misa,han2021improving}. However, we argue that each modality contributes differently and thus should be assigned a dynamic weight. To this end, we design a simple yet effective weighted feature fusion network.

Given that static weights may not effectively capture the nuances of different multimodal contexts, we introduce learning dynamic weights adaptive to these contexts. Specifically, we employ three distinct neural networks, one for each modality, to compute the modality-specific importance score $s^\mathrm{M} \in \mathbb{R}$:
\begin{align}
    s^{\mathrm{M}} = \mathbf{W}_{2}^{\mathrm{M}}
    \bigl(\operatorname{Dropout}[\operatorname{ReLU}(\mathbf{W}_{1}^{\mathrm{M}}(\mathbf{x}^{\mathrm{M}}))]\bigr),
\end{align}
where $\mathbf{W}_{1}^{\mathrm{M}} \in \mathbb{R}^{D_{\mathrm{M}} \times H_{\mathrm{w}}}$ and $\mathbf{W}_{2}^{\mathrm{M}} \in \mathbb{R}^{H_{\mathrm{w}} \times 1}$ are learnable weights, and $H_{\mathrm{w}}$ is the size of the hidden layer. We then normalize each score via the softmax function:
\begin{align}
    w^{\mathrm{M}} = \frac{\exp(s^{\mathrm{M}})}{\sum_{\mathrm{M} \in \{\mathrm{T, A, V}\}} \exp(s^{\mathrm{M}})},
\end{align}
yielding the weight $w^{\mathrm{M}} \in (0, 1)$. The final fused representation $\mathbf{z}_{\textrm{F}}$ is computed as
\begin{align}
    \mathbf{z}_{\textrm{F}} = \sum_{\mathrm{M} \in \{\mathrm{T, A, V}\}} w^{\mathrm{M}} \cdot \mathbf{x}^{\mathrm{M}}.
\end{align}

The proposed fusion network introduces a novel approach by dynamically learning modality weights through neural networks, which contrasts with traditional fusion strategies that rely on simple methods such as vector concatenation or addition. Unlike recent transformer-based methods, such as cross-modal attention mechanisms~\cite{tsai2019multimodal, huang2024sdif} or those conditioned on semantic shifts in non-verbal modalities~\cite{rahman2020integrating}, these approaches often overlook the varying importance of different modalities or treat one modality (e.g., text) as dominant, with the others serving as secondary. This can lead to suboptimal fusion, where the contributions of each modality are not appropriately balanced. In contrast, our method enhances fusion by adaptively weighting each modality based on its contextual relevance, resulting in a more effective integration of multimodal data. As demonstrated in Section~\ref{ablation_studies}, this approach significantly outperforms existing fusion techniques in terms of both flexibility and performance.

\subsection{Multimodal Representation Learning}
To learn robust representations for both ID classification and OOD detection, we aim to improve the discrimination ability of each multimodal sample among ID and pseudo-OOD data at both coarse-grained and fine-grained levels. 

Coarse-grained distinction primarily focuses on binary classification tasks (i.e., distinguishing between ID and OOD). This stage aims to capture more global and general characteristics of the data~\cite{yang2024generalized}. The model first learns to identify modality-invariant differences between ID and OOD samples, which is crucial for establishing an initial decision boundary in the multimodal feature space. This coarse distinction provides the foundation for further multimodal representation learning, where more complex interactions can be explored.

\textbf{Binary Training.} To grasp these fundamental distinctions, we perform binary classification using both ID samples and our generated pseudo-OOD samples. Specifically, we introduce the binary classifier $\Phi_{\textrm{b}} \in  \mathbb R^{D_{\mathrm{H}} \times 2}$, which consists of a stack of two non-linear layers. The $\operatorname{Softmax}$ operation is applied to the classifier outputs to obtain the prediction probabilities $\text{P}_{\textrm{b}}=\operatorname{Softmax}(\Phi_{\textrm{b}}(\mathbf{z}_{\textrm{F}}))$, serving as confidence scores. We use binary cross-entropy to define the loss in the coarse-grained training stage $\mathcal{L}_{\textrm{coarse}}$:
\begin{align}
    \mathcal{L}_{\textrm{coarse} } &= -\frac{1}{B}\sum_{i=1}^B\sum_{c \in \{0,1\}}[ y_{\textrm{b}, i}^c \log(\text{P}^c_{\textrm{b}, i})],
\end{align} where $B$ is sample count in a mini-batch, and $y_{\textrm{b}}$ denotes the binary labels with 0 indicating OOD and 1 indicating ID. 

Fine-grained distinction then refines the model’s discriminative power, allowing it to capture subtle differences within ID classes or instance-level deviations between ID and OOD. This goes beyond merely separating ID from OOD, targeting more nuanced features in the multimodal space~\cite{10097558,wang2023app}. In this stage, we apply multi-class training and contrastive learning, enabling the model to capture intricate cross-modal relationships and improve its sensitivity to these subtle differences.

\textbf{Multi-class Training}. After mastering the basic binary distinguishing characteristics, we proceed to learning fine-grained discriminative information, focusing on capturing inter-class separation properties within ID data through multi-class classification. Since multimodal representations are obtained by weighted summation of features from different modalities, the scale of each modality’s feature norm may vary and is often susceptible to noise. Therefore, we replace the conventional linear classifier with a cosine classifier~\cite{qi2018low}, which, by L2-normalizing both the representation and the weight vector, focuses on directional information. Specifically, the cosine classifier $\Phi_{\text{cos}} \in \mathbb{R}^{D_{\mathrm{H}} \times K}$ operates directly on the fused multimodal representations $\mathbf{z}_{\textrm{F}}$ as follows:
\begin{align}
\Phi_{\text{cos}}(\mathbf{z}_{\textrm{F}})=\gamma \cdot \frac{\mathbf{z}_{\textrm{F}}}{\|\mathbf{z}_{\textrm{F}}\|_2}
 \cdot \frac{\textbf{W}_{\textrm{cos}}^\top}{\|\textbf{W}_{\textrm{cos}}\|_2},
\end{align} where $K$ is the number of ID classes, $\gamma$ is a scaling factor that controls the output logits within a proper range, and $|\cdot|2$ denotes the L2 norm. This classifier computes the cosine similarities between the fused representation $\mathbf{z}{\textrm{F}}$ and the weight matrix $\textbf{W}_{\textrm{cos}}$ by first performing L2 normalization on each and then computing the dot product. This process enhances the model’s discriminative ability across different known classes and effectively promotes the complementary directional information across modalities. Our experiments demonstrate that the cosine classifier generally outperforms the conventional linear layer in both ID classification and OOD detection (referred to as $\textit{w / o Cosine}$ in Table~\ref{ablation_results}).

We train the cosine classifier with the softmax loss under the supervision of multi-class labels from the ID data:
\begin{align}
      \mathcal{L}_{\textrm{m}} = -\frac{1}{B}\sum_{i=1}^B\sum_{c \in \mathcal{Y}}[ y_{\textrm{m}, i}^c \log(\operatorname{Softmax}(\Phi_{\text{cos}}({\mathbf{z}_{\textrm{F}}}))^c)],
\end{align}
where $\mathcal{Y} \in \{0, 1, \cdots, K-1\}$ denotes the ID label set.  

\textbf{Contrastive Learning}. In addition to capturing fine-grained inter-class separation among ID data, we explore instance-level interactions between ID and OOD data, which are crucial for enhancing the model's generalization ability across both tasks. To leverage these insights, we apply contrastive learning to both ID and the generated OOD data. 

We define samples with identical labels within ID data or each sample with its augmentations as positive pairs, while all other combinations are considered negative pairs. Each piece of OOD data is treated as negative when paired with other samples because pseudo-OOD samples are convex combinations of multimodal representations from various ID classes, exhibiting significant semantic differences in the feature space. The contrastive losses are defined as follows:
\begin{align}
    &\mathcal{L}_{\textrm{cl-ID}}= \nonumber \\
   &-\frac{1}{2B} \sum_{i \in \mathcal{B}_{\textrm{ID}}} \frac{1}{|\mathcal{P}(i)|} \sum_{p \in \mathcal{P}(i)} \log \frac{\exp(\textrm{sim}(\boldsymbol{l}_{i}, \boldsymbol{l}_{p}) / \tau)}{\sum_{k=1}^{2B} \mathbb{I}_{[k \neq i]} \exp(\textrm{sim}(\boldsymbol{l}_{i}, \boldsymbol{l}_{k}) / \tau)},\\
   &\mathcal{L}_{\textrm{cl-OOD}}= -\frac{1}{2B} \sum_{j \in \mathcal{B}_{\textrm{OOD}}} \log \frac{\exp(\textrm{sim}(\boldsymbol{l}_{j}, \tilde{\boldsymbol{l}}_{j}) / \tau)}{\sum_{k=1}^{2B} \mathbb{I}_{[k \neq j]} \exp(\textrm{sim}(\boldsymbol{l}_{j}, \boldsymbol{l}_{k}) / \tau)},
\end{align}
where $\mathcal{B}_{\textrm{ID}}$ and $\mathcal{B}_{\textrm{OOD}}$ denote the sets of indices for ID and OOD data, respectively. $\mathcal{P}(i)$ refers to the indices of the samples sharing the same label as $i$. $\boldsymbol{l}$ represents the contrastive feature for each sample, derived from $\mathbf{z}_{\textrm{F}}$ through a linear layer tailored for contrastive learning, denoted as $\boldsymbol{l} = \Phi_{\text{cl}}(\mathbf{z}_{\textrm{F}})$. $\textrm{sim}(\boldsymbol{l}_{a}, \boldsymbol{l}_{b})$ denotes the dot product between two L2-normalized vectors. $\mathbb{I}_{[\textrm{condition}]}$ is an indicator function that outputs 1 if the condition is met, otherwise 0. $\tau$ represents the temperature hyper-parameter. Each mini-batch includes an equal number of ID and OOD samples, totaling $B$ samples, and uses $\textit{dropout}$~\cite{gao2021simcse} to generate a positive augmented sample $\tilde{\boldsymbol{l}}$ for each original one, resulting in 2$B$ samples per mini-batch. The final contrastive loss,  $\mathcal{L}_{\textrm{cl}}$, is defined as $\mathcal{L}_{\textrm{cl}} = \mathcal{L}_{\textrm{cl-ID}} + \mathcal{L}_{\textrm{cl-OOD}}$. Through $\mathcal{L}_{\textrm{cl}}$, the model learns fine-grained relations among ID and OOD samples by pulling positive pairs closer and pushing negative pairs further apart. This enhances the model's ability to discriminate between both ID and OOD data. Finally, we jointly optimize the multi-class training and contrastive learning losses during the fine-grained training stage:  $\mathcal{L}_{\textrm{fine}} = \mathcal{L}_{\textrm{m}} + \mathcal{L}_{\textrm{cl}}$. 

This multi-granularity optimization introduces a hierarchical scheme that facilitates the learning process by progressively tackling tasks of increasing difficulty. Focusing too much on coarse-grained tasks may overlook important nuanced distinctions, while overemphasizing fine-grained tasks could lead to overfitting or inefficiency. Our multi-stage approach addresses this by first focusing on coarse-grained learning to establish a robust multimodal decision boundary between ID and OOD data. Based on this, the fine-grained objectives further refine the model’s ability to distinguish nuanced class- and instance-level multimodal variations. This method effectively balances both coarse- and fine-grained distinctions, improving adaptation to challenging multimodal samples and ensuring robustness in both ID classification and OOD detection.

\subsection{Inference} \label{mahalanobis}

For ID classification, we obtain logits from the cosine classifier and select the class with the highest probability as the predicted label. For OOD detection, we employ the Mahalanobis distance~\cite{lee2018simple} as the scoring function on the fused representations to generate discriminative scores:
\begin{align}
    &\text{Score}_{\text{Maha.}} 
    = 
    \min_k \left\{ 
         (\mathbf{z}_{\textrm{F}}^{\textrm{Test}} - \boldsymbol{\mu}_k)^\top 
         \hat{\Sigma}_k^{-1}
         (\mathbf{z}_{\textrm{F}}^{\textrm{Test}} - \boldsymbol{\mu}_k) 
       \right\}, \\
    &\boldsymbol{\mu}_k 
    = 
    \frac{\sum_{y_i=k}\mathbf{z}_{\textrm{F}, i}^{\textrm{Train}}}{N_k},
    \hat{\Sigma}_k 
    = 
    \frac{\sum_{y_i=k}
    ( \mathbf{z}_{\textrm{F}, i}^{\textrm{Train}} - \boldsymbol{\mu}_k)
    ( \mathbf{z}_{\textrm{F}, i}^{\textrm{Train}} - \boldsymbol{\mu}_k)^\top}{N_k-1},
    \label{ma_score}
\end{align}
where $\mathbf{z}_{\textrm{F}}^{\textrm{Train}}$ and $\mathbf{z}_{\textrm{F}}^{\textrm{Test}}$ denote the multimodal features extracted from the training and testing sets, respectively. $\boldsymbol{\mu}_k$ represents the mean of the training samples from class $k$, $N_k$ corresponds to the number of samples in class $k$, and $\hat{\Sigma}_k$ is the covariance matrix computed from the training samples.

In our experiments, all compared methods use this OOD detection approach to ensure a fair comparison. The Mahalanobis distance demonstrates robust performance across all datasets, and a detailed comparison with other state-of-the-art OOD detection methods is presented in Section~\ref{OOD_comparison}.

\begin{table*}[t!]
    \caption{ \label{datasets} Statistics of the MIntRec, MELD-DA, and IEMOCAP-DA datasets. \#C, \#Video, and \#U denote the number of intent classes, videos, and utterances, respectively. \#Train, \#Valid, and \#Test indicate the number of utterances in the training, validation, and testing sets. OOD utterances only appear in the testing set. Max. and Avg. represent the maximum and average lengths.}
    \centering
    \resizebox{2\columnwidth}{!}{
    \begin{tabular}{@{} llccccccccc @{}}
        \toprule
        Datasets & \#C  & \#Video & \#U & \#Train & \#Valid & \makecell{\#Test \\(ID / OOD)} & \makecell{Text Length \\(Max. / Avg.)} & \makecell{Video Length \\(Max. / Avg.)} & \makecell{Audio Length \\(Max. / Avg.)} & \makecell{Video \\Hours}\\
        \midrule
        MIntRec & 20 & 43 & 2,674  & 1,334 & 445 & 445 / 450 & 30 / 7.04 & 230 / 54.37  & 480 / 118.03 & 1.47 (h) \\ 
        MELD-DA & 11 & 1039 & 9,988  & 6,561 & 937 & 1,875 / 615 & 70 / 7.95 & 250 /  72.99 & 530 / 152.43 & 8.75 (h)\\
        IEMOCAP-DA  & 11 & 302 & 9,416 & 6,541 & 935 & 1,870 / 70 & 44 / 11.53 &  230 / 133.77& 380 / 223.63 & 11.68 (h)\\
        \bottomrule
    \end{tabular}
    }
\end{table*}
\section{Experiments}
\subsection{Datasets}
\subsubsection{\textbf{Multimodal Intent Dataset}}
The MIntRec dataset~\cite{zhang2022mintrec} provides a pioneering resource in this area, containing 2,224 multimodal utterances across various modalities (i.e., text, video, audio) with high-quality annotations spanning 20 fine-grained intent categories in real-world conversational scenarios. The original dataset serves as the ID data, and due to its lack of OOD data, we have manually constructed an OOD dataset, which is described in detail below.

The motivation for constructing the OOD data is that real-world scenarios often include unexpected out-of-scope utterances that fall outside the well-annotated known classes during dialogue interactions. Such data, if not properly handled, can degrade the performance of dialogue systems. To address this challenge and mitigate data scarcity, we construct the OOD dataset through the following two key steps: 

\textbf{Data Preparation}: As recommended in~\cite{zhang2022mintrec}, we begin by collecting raw video clips from the TV series \textit{Superstore}, ensuring that the segments selected are distinct from those in the original dataset. These segments are chosen to reflect a variety of conversational contexts, including different settings, characters, and emotional tones, thereby ensuring that the OOD data captures a broad range of potential out-of-scope scenarios. The raw video clips are then split into shorter clips based on timestamps and subtitles. For each clip, we extract the corresponding audio segments, which are then paired with the subtitles to form the multimodal data.

\textbf{Data Annotation}: 
We employ five annotators, each trained to carefully review the multimodal content of each sample. Before beginning the annotation process, the annotators familiarize themselves with the definitions of the 20 intent categories and review a series of typical examples. Each annotator then labels whether a sample is an ID or OOD instance based on the textual, audio, and video information. To ensure reliability and consistency, only those samples for which all five annotators agree on the classification (i.e., all label the sample as OOD) are selected. In total, 450 utterances are chosen to comprise the OOD data for the MIntRec dataset.

\subsubsection{\textbf{Multimodal Dialogue Act Datasets}} 
As dialogue acts represent coarse-grained communicative intents during conversational interactions, we utilize two multimodal dialogue datasets, MELD-DA and IEMOCAP-DA. MELD-DA, derived from the MELD dataset~\cite{poria2019meld}, originates from the TV series \textit{Friends} and contains over 13,000 utterances from 1,433 dialogues. IEMOCAP-DA is based on the IEMOCAP dataset~\cite{busso2008iemocap}, which includes videos from both scripted plays and spontaneous interactions involving 10 actors, totaling nearly 12 hours. For dialogue act labels, we adopt the well-annotated labels from the EMOTyDA dataset~\cite{saha-etal-2020-towards}. The employed dialogue act taxonomy includes 12 frequently occurring tags derived from the Switchboard corpus~\cite{stolcke2000dialogue}, comprising 11 known classes, namely: \textit{Acknowledge} (a), \textit{Agreement} (ag), \textit{Answer} (ans), \textit{Apology} (ap), \textit{Backchannel} (b), \textit{Command} (c), \textit{Disagreement} (dag), \textit{Greeting} (g), \textit{Statement-Opinion} (o), \textit{Question} (q), \textit{Statement-Non-Opinion} (s), and one \textit{Others} (oth) class, which encompasses all remaining dialogue act types. Therefore, we treat samples within the known classes as ID data and those with the \textit{Others} tag as OOD data. Following~\cite{hendrycks17baseline,liangenhancing}, OOD data only appear in the testing set during OOD detection. Detailed statistics and data splits are provided in Table~\ref{datasets}.

\subsection{Baselines}
\label{baselines}
Due to the absence of multimodal methods specifically designed for OOD detection, we establish baselines using state-of-the-art multimodal fusion methods. These include TEXT, MulT~\cite{tsai2019multimodal}, MAG-BERT~\cite{rahman2020integrating}, MMIM~\cite{han2021improving}, SDIF~\cite{huang2024sdif}, SPECTRA~\cite{yu2023speech}, TCL-MAP~\cite{zhou2024token}, as well as MIntOOD (R) and MIntOOD (T), two variants of our method. Detailed descriptions of these methods are provided below:  

\textbf{TEXT:} This baseline utilizes only the text modality. We fine-tune the pre-trained BERT language model with a classifier for intent classification, supervised by the training targets.

\textbf{MulT:} It first employs six cross-modal transformers to capture interactions between modality pairs. Then, for each transformer that includes the target modality, a self-attention mechanism is applied to the concatenated outputs to capture temporal information.

\textbf{MAG-BERT:} It introduces a plug-and-play module that integrates non-verbal inputs between layers of pre-trained transformers. This module computes the displacement of non-verbal vectors and performs a weighted summation with verbal displacement vectors to produce a multimodal representation.

\textbf{MMIM:} It learns modality-invariant information by optimizing mutual information bounds between text and other modalities. It computes inverse correlations between multimodal and unimodal features, optimized via noise contrastive estimation~\cite{oord2018representation}. Entropy estimators are not used as they adhere to sentiment polarities, which are not applicable to our task.

\textbf{SPECTRA:} It comprises a text encoder, an audio encoder, and a fusion module, pre-trained on a temporal position prediction task for explicit text-audio alignment and fine-tuned for enhanced multimodal intent recognition.

\textbf{TCL-MAP:} It first generates a modality-aware prompt to refine the text modality and then uses the ground truth label to create an augmented sample. Next, a contrastive learning mechanism is applied to the token to leverage ground-truth semantics and guide the learning of non-verbal modalities.

\textbf{SDIF:} It employs a text-centric shallow interaction module to align video and audio features with text, followed by a transformer-based deep interaction module to refine cross-modal fusion.

\textbf{MIntOOD (R):} A variant of MIntOOD that utilizes real OOD data for training and validation. For a fair comparison, the OOD data from the MELD-DA dataset is used for the MIntRec dataset, while the OOD data from the MIntRec dataset is used for the MELD-DA and IEMOCAP-DA datasets.

\textbf{MIntOOD (T):} A variant of MIntOOD, differing mainly in that it uses sentence-level representations from BERT rather than multimodal fused representations.

\begin{table*}[t!]\scriptsize
\centering
\caption{ID classification and OOD detection results on the MIntRec, MELD-DA, and IEMOCAP-DA datasets.}
\label{main_results}
\resizebox{18cm}{!}{
\begin{tabular}
{@{\extracolsep{2pt}}cl|cccccc|ccccc}
    \toprule
   \multirow{2}{*}{Datasets} &  \multirow{2}{*}{Methods}& \multicolumn{6}{c|}{ID Classification} & \multicolumn{5}{c}{OOD Detection}\\
   
    &&\makecell{ACC ($\uparrow$)} 
    & \makecell{WF1 ($\uparrow$)} 
    & \makecell{WP ($\uparrow$)} 
    & \makecell{F1 ($\uparrow$)} 
    & \makecell{P ($\uparrow$)}
    & \makecell{R ($\uparrow$)}
    & \makecell{DER ($\downarrow$)} 
    &\makecell{FPR95 ($\downarrow$)} 
    & \makecell{AUPR  In ($\uparrow$)} 
    & \makecell{AUPR Out ($\uparrow$)} 
    & \makecell{AUROC ($\uparrow$)}\\
    \midrule
   \multirow{10}{*}{\rotatebox{90}{MIntRec}} 
   & TEXT & 70.34 & 70.46 & 71.12 & 67.06 & 67.53 & 67.24 & 40.25 & 76.00 & 74.52 & 75.14 & 76.23 \\
   & MAG-BERT & 72.00 & 71.78 & 72.45 & 68.36 & 69.01 & 68.92 & 41.80 & 79.07 & 74.05 & 71.31 & 74.48 \\
   & MulT & 72.31 & 72.07 & 72.24 & 68.97 & 69.73 & 68.83 & 42.13 & 79.78 & 74.77 & 71.45 & 75.21 \\
   & MMIM & 72.05 & 71.97 & 72.80 & \underline{69.68} & \underline{70.59} & 69.81 & 40.45 & 76.45 & 75.84 & 73.89 & 75.85\\
   & SPECTRA & 71.01 & 70.83 & 71.90 & 67.87 & 69.41 & 68.10 & 40.78 & 77.02 & 74.88 & 73.92 & 75.35 \\
   & TCL-MAP & \underline{73.21} & \underline{72.73} & \underline{73.02} & 69.02 & 69.39 & \underline{69.88} & 41.40 & 78.31 & 74.71 & 73.19 & 75.36 \\
   & SDIF & 71.42 & 71.24 & 71.82 & 68.53 & 70.33 & 67.86 & 39.77 & 75.42 & 72.47 & 73.06 & 73.40 \\
   \cmidrule{2-13}
   & MIntOOD (R) & 72.18 & 71.86 & 72.59 & 68.29 & 69.75 & 68.43 & 40.08 & 75.60 & \underline{77.67} & 75.35 & 78.55 \\  
   & MIntOOD (T) & 72.14 & 71.78 & 72.00 & 68.30 & 68.88 & 68.45 & \underline{37.44} & \underline{70.31} & 76.57 & \underline{78.38} & \underline{78.95}\\
   & MIntOOD & \textbf{74.34} & \textbf{74.15} & \textbf{74.51} & \textbf{70.94} & \textbf{72.24} & \textbf{70.46} &  \textbf{36.03} &  \textbf{67.56} &  \textbf{79.69} &  \textbf{80.09} &  \textbf{80.54}\\
   \midrule
   \midrule
   \multirow{10}{*}{\rotatebox{90}{MELD-DA}}
   & TEXT & 63.54 & 62.59 & 62.64 & 54.48 & 57.07 & 53.55 & 59.67 & 77.63 & 84.57 & 44.55 & 67.38\\
   & MAG-BERT & \underline{64.50} & 63.16 & 63.14 & 54.30 & 58.81 & 53.51 & 58.07 & 75.54 & 86.63 & 49.02 & 72.32\\
   & MulT & 63.35 & 62.28 & 62.96 & 54.20 & 58.45 & 53.57 & 63.03 & 82.08 & 86.25 & 40.82 & 69.62\\
   & MMIM & 64.45 & 63.08 & 63.57 & 55.27 & 59.88 & 54.02 & \underline{56.03} & \underline{72.78} & 87.74 & \underline{51.59} & 73.83\\
   & SPECTRA & 60.81 & 60.31 & 60.93 & 54.62 & 55.66 & \textbf{55.28} & 63.57 & 82.83 & 83.68 & 40.36 & 64.62 \\
   & TCL-MAP & 64.23 & 62.94 & 62.73 & 53.98 & 57.10 & 53.22 & 58.48 & 76.03 & 86.17 & 48.82 & 71.79 \\
   & SDIF & 64.33 & \underline{63.19} & \underline{63.75} & \underline{55.56} & 62.11 & 54.00 & 56.81 & 73.82 & \underline{87.87} & 49.73 & 74.46 \\
   \cmidrule{2-13}
   & MIntOOD (R) & \underline{64.50} & 62.45 & 62.80 & 53.65 & 59.90 & 52.34 & 57.36 & 75.60 & 87.47 & 51.08 & \underline{74.57} \\  
   & MIntOOD (T) & 64.32 & 62.61 & 63.58 & 55.16 & \underline{64.71} & 53.33 & 57.26 & 74.37 & 86.88 & 50.49 & 73.40 \\
   & MIntOOD & \textbf{65.00} & \textbf{63.53} & \textbf{64.62} & \textbf{56.20} & \textbf{65.09} & \underline{54.20} &  \textbf{55.54} &  \textbf{72.13} &  \textbf{88.85} &  \textbf{54.45} &  \textbf{76.95}\\
   \midrule
   \midrule
   \multirow{10}{*}{\rotatebox{90}{IEMOCAP-DA}}
   & TEXT & 73.97 & 73.88 & 74.53 & 70.28 & 70.18 & \underline{72.00} & 57.67 & 59.71 & 99.45 & 22.54 & 87.92\\
   & MAG-BERT & 74.52 & 74.33 & 74.64 & 70.80 & 71.70 & 71.59 & 87.38 & 90.57 & 99.30 & 14.03 & 84.30\\
   & MulT & 73.87 & 73.73 & 73.97 & 70.75 & 72.01 & 70.25 & 90.68 & 94.00 & 99.04 & 7.97 & 77.69\\
   & MMIM & 74.48 & 74.32 & 74.69 & 71.92 & 73.08 & 71.83 & 71.18 & 73.72 & 99.09 & 15.45 & 79.94 \\
   & SPECTRA & 74.54 & 74.43 & \underline{74.93} & 70.35 & 72.17 & 70.10 & 93.42 & 96.86 & 99.23 & 10.20 & 81.47 \\
   & TCL-MAP & \underline{74.61} & \underline{74.44} & 74.72 & 72.15 & \underline{74.37} & 70.98 & 95.87 & 99.43 & 98.77 & 6.61 & 71.72 \\
   & SDIF & 74.49 & 74.38 & 74.81 & \underline{72.18} & \textbf{74.64} & 70.89 & 64.33 & 66.57 & 99.55 & 18.94 & 88.83 \\
   \cmidrule{2-13}
   & MIntOOD (R) & 74.47 & 71.86 & 72.59 & 68.29 & 69.75 & 68.43 & \underline{14.19} & \underline{14.57} & \underline{99.85} & \textbf{85.08} & \underline{96.52} \\  
   & MIntOOD (T) & 74.46 & 74.36 & 74.73 & 71.50 & 73.19 & 70.75 & 29.87 & 30.86 & 99.76 & 40.13 & 94.63 \\
   & MIntOOD & \textbf{74.88} & \textbf{74.77} & \textbf{75.24} & \textbf{72.82} & 74.28 & \textbf{72.42} & \textbf{12.82} & \textbf{13.14} & \textbf{99.88} & \underline{84.03} & \textbf{97.19} \\
    \bottomrule 
\end{tabular}}
\end{table*}

\subsection{Evaluation Metrics}
\textbf{ID Classification}: To evaluate the performance of ID classification, we use six metrics: accuracy (ACC), weighted F1-score (WF1), weighted precision (WP), F1-score (F1), precision (P), and recall (R). These metrics are defined as follows:
\begin{align}
&\text{P}=\frac{1}{K}\sum_{i=1}^{K} \frac{\text{TP}_{C_{i}}}{\text{TP}_{C_{i}}+\text{FP}_{C_{i}}}, \text{R}=\frac{1}{K}\sum_{i=1}^{K} \frac{\text{TP}_{C_{i}}}{\text{TP}_{C_{i}}+\text{FN}_{C_{i}}}, \\
&\text{ACC}=\sum_{i=1}^{T}\mathbb{I}({y^{\text{Pred}}=y^{\text{GT}}}), \text{F1} = 2 \times \frac{\text{P} \times \text{R}}{\text{P} + \text{R}}, \\
&\text{WP}=\frac{1}{K}\sum_{i=1}^{K}\frac{T_{C_i}}{T}\times \frac{\text{TP}_{C_{i}}}{\text{TP}_{C_{i}}+\text{FP}_{C_{i}}}, \\
&\text{WF1}=\frac{1}{K}\sum_{i=1}^{K}\frac{T_{C_i}}{T}\times \frac{2\times\text{P}_{C_{i}}\times\text{R}_{C_{i}}}{\text{P}_{C_{i}}+\text{R}_{C_{i}}},
\end{align}
where $K$ is the number of classes, $C_i$ is the $i^{\text{th}}$ class, and $\text{TP}, \text{FP}, \text{FN}$ denote the true positives, false positives, and false negatives, respectively. Here, $y^{\text{Pred}}$ and $y^{\text{GT}}$ denote the predicted and ground truth labels, $T$ is the total number of ID samples in the testing set, and $T_{C_i}$ is the number of ID samples from class $C_i$ in the testing set.

\textbf{OOD Detection}: Following~\cite{liangenhancing,ryu2018out,lee2018simple}, we evaluate OOD detection performance using five typical metrics: DER, FPR95, AUPR-In, AUPR-Out, and AUROC. These metrics are selected to comprehensively assess OOD detection performance under different conditions. Lower values are better for the first two metrics, while higher values are better for the latter three. The details of these metrics are as follows:

\begin{itemize}
    \item \textbf{FPR95}: Defined as the false positive rate when the true positive rate (TPR) is at least 95\%, it represents the probability of an OOD sample being misclassified as ID. Here, TPR is defined as: TP / (TP+FN).
    \item \textbf{DER}: Calculated as $P_{\text{ID}}\cdot(1-\text{TPR})+P_{\text{OOD}}\cdot\text{FPR}$ when TPR is 95\%. We assume equal appearance probabilities for ID and OOD samples during testing, as suggested in~\cite{liangenhancing}. Here, FPR is defined as: FP / (TN + FP), where TN denotes the number of true negatives.
    \item \textbf{AUPR}: The area under the precision-recall curve, with AUPR-In and AUPR-Out treating ID and OOD samples as positives, respectively. 
    \item  \textbf{AUROC}: The area under the receiver operating characteristic curve, illustrating the relationship between TPR and FPR at various thresholds. 
\end{itemize}
These five metrics together offer a robust evaluation of the model’s OOD detection capabilities across various thresholds, balancing the trade-off between rejecting OOD samples and correctly classifying ID samples.

\subsection{Experimental Settings}
\label{experimental_settings}

In the experiments, ID data presents in the training, validation, and testing sets, while OOD data only appears in the testing set. Each method is tuned using the ID data from the validation set, and the best model is saved based on an early stopping strategy.  During inference, we evaluate ID classification performance on the ID data and generate Mahalanobis scores (as defined in Eq.~\ref{ma_score}) for both ID and OOD data to assess OOD detection performance.

For feature extraction, sequence lengths $L_{\textrm{T}}$, $L_{\textrm{V}}$, and $L_{\textrm{A}}$ are set to (30, 230, 480), (70, 250, 530), and (44, 230, 380) for the MIntRec, MELD-DA, and IEMOCAP-DA datasets, respectively. Feature dimensions $D_{\textrm{T}}$, $D_{\textrm{V}}$, and $D_{\textrm{A}}$ are set at 768, 1024, and 768, respectively. During OOD data generation, the number of selected embedded examples $k$ is set to 3 for all three datasets. The sampling parameter $\alpha$ of the Dirichlet distribution is set to (2, 0.7, 0.7). The hidden feature dimension $D_{\mathrm{H}}$ is set to 768, and the hidden size of the weighted feature fusion network $H_w$ is 256. The scaling factor $\gamma$ in the cosine classifier is set to (16, 16, 32), and the temperature $\tau$ for contrastive learning is set to (2, 1, 0.7). The number of training epochs is 100, and the batch size is 32. We utilize the PyTorch library implemented in HuggingFace~\cite{wolf2020transformers} for the pre-trained BERT language model, adopting the AdamW~\cite{loshchilov2018decoupled} optimizer with learning rates of (3e-5, 4e-6, 3e-6). For a fair comparison, all experimental results are reported as averages over five runs with random seeds ranging from 0 to 4. The experiments are conducted on an NVIDIA Tesla V100-SXM2.

\begin{table*}[t!]\small
\centering
\caption{Ablation studies on the MIntRec, MELD-DA, and IEMOCAP-DA datasets.}
\label{ablation_results}
\resizebox{18cm}{!}{
\begin{tabular}
{@{\extracolsep{2pt}}cl|cccccc|ccccc}
    \toprule
   \multirow{2}{*}{Datasets} &  \multirow{2}{*}{Methods}& \multicolumn{6}{c|}{ID Classification} & \multicolumn{5}{c}{OOD Detection}\\
   
    &&\makecell{ACC ($\uparrow$)} 
    & \makecell{WF1 ($\uparrow$)} 
    & \makecell{WP ($\uparrow$)} 
    & \makecell{F1 ($\uparrow$)} 
    & \makecell{P ($\uparrow$)}
    & \makecell{R ($\uparrow$)}
    & \makecell{DER ($\downarrow$)} 
    &\makecell{FPR95 ($\downarrow$)} 
    & \makecell{AUPR  In ($\uparrow$)} 
    & \makecell{AUPR Out ($\uparrow$)} 
    & \makecell{AUROC ($\uparrow$)}\\
    \midrule
   \multirow{9}{*}{\rotatebox{90}{MIntRec}} 
  &  Fusion (Add)  & 73.04 & 72.71 & 73.21 & 68.64 & 70.09 & 68.43 & 38.39 & 72.45 & 77.84 & 78.43 & 78.78 \\
    &  Fusion (Concat) & 72.90 & 72.67 & 72.95 & 68.91 & 69.94 & 68.66 & \underline{38.90} & 73.29 & 78.07 & 77.57 & 79.06\\
    &  Fusion (MulT) & 71.19 & 70.75 & 71.75 & 67.64 & 69.34 & 67.82 & 40.85 & 77.16 & 76.11 & 74.15 & 76.46\\
    &  Fusion (MAG-BERT) & 72.27 & 71.92 & 72.87 & 68.44 & 69.36 & \underline{69.14} & 39.12 & 73.78 & 76.41 & 76.17 & 78.12\\
    &  Fusion (SDIF) & 71.82 & 71.47 & 71.67 & 68.70 & 70.03 & 68.19 & 39.45 & 74.36 & 77.12 & 76.46 & 77.64\\
    & w / o Contrast  & 72.90 & 72.74 & 73.32 & 69.38 & 70.69 & 69.12 & 37.93 & \underline{71.42} & \underline{78.84} & \underline{78.90} & \underline{79.83}\\
    & w / o Cosine & 68.90 & 68.19 & 70.34 & 64.25 & 67.12 & 65.07 & 39.70 & 75.02 & 72.84 & 75.08 & 75.39\\
    & w / o Binary & \underline{73.08} & \underline{72.89} & \underline{73.40} & \underline{69.42} & \underline{70.95} & 68.94 & 39.57 & 74.71 & 78.34 & 77.06 & 78.44\\ 
    & Full & \textbf{74.34} & \textbf{74.15} & \textbf{74.51} & \textbf{70.94} & \textbf{72.24} & \textbf{70.46} &  \textbf{36.03} &  \textbf{67.56} &  \textbf{79.69} &  \textbf{80.09} &  \textbf{80.54}\\
   \midrule
   \midrule
   \multirow{9}{*}{\rotatebox{90}{MELD-DA}}
   & Fusion (Add) & 63.22 & 61.01 & 61.90 & 52.36 & 58.06 & 50.80 & 56.92 & 73.98 & \underline{88.31} & \underline{51.42} & \underline{75.23}\\
    & Fusion (Concat) & 63.65 & 61.34 & 62.63 & 52.75 & 61.50 & 51.19 & 59.85 & 77.89 & 87.19 & 48.11 & 73.17\\
    &  Fusion (MulT) & 63.61 & 61.25 & 62.64 & 51.83 & 61.72 & 50.43 & 59.48 & 77.37 & 87.49 & 45.18 & 73.43\\
    &  Fusion (MAG-BERT) & \underline{64.44} & 62.01 & 62.62 & 53.01 & 58.42 & 52.18 & \underline{55.66} & \underline{72.29} & 87.59 & 50.90 & 74.08\\
    &  Fusion (SDIF) & 63.47 & \underline{62.23} & 62.62 & \underline{54.56} & 59.16 & 53.16 & 59.10 & 76.88 & 86.27 & 47.16 & 71.37\\
    & w / o Contrast & 64.13 & 62.12 & \underline{63.01} & 53.95 & \underline{61.87} & 52.45 & 56.95 & 74.02 & 86.67 & 50.25 & 73.27\\
    & w / o Cosine & 63.00 & 60.58 & 62.24 & 53.63 & 59.22 & \underline{53.21} & 61.83 & 80.49 & 86.47 & 46.54 & 72.47\\
    & w / o Binary & 63.50 & 60.51 & 62.01 & 51.18 & 59.24 & 49.94 & 58.76 & 76.42 & 87.61 & 49.97 & 74.32\\ 
    & Full & \textbf{65.00} & \textbf{63.53} & \textbf{64.62} & \textbf{56.20} & \textbf{65.09} & \textbf{54.20} &  \textbf{55.54} &  \textbf{72.13} &  \textbf{88.85} &  \textbf{54.45} &  \textbf{76.95}\\
   \midrule
   \midrule
   \multirow{9}{*}{\rotatebox{90}{IEMOCAP-DA}}
& Fusion (Add) & 74.57 & 74.44 & 74.83 & 71.14 & 72.21 & 70.99 & 21.92 & 22.57 & 99.81 & 38.75 & 95.14 \\
& Fusion (Concat) & 74.03 & 73.90 & 74.08 & 70.16 & 72.28 & 69.52 & 43.13 & 44.57 & 99.34 & 26.82 & 87.72 \\
&  Fusion (MulT) & 74.39 & 74.20 & 74.47 & 72.20 & 73.01 & 72.32 & 95.41 & 98.86 & 99.18 & 9.17 & 80.67\\
    &  Fusion (MAG-BERT) & 74.16 & 74.02 & 74.36 & 71.52 & 72.78 & 71.12 & 64.82 & 67.14 & 99.18 & 30.82 & 82.54\\
    &  Fusion (SDIF) & 74.78 & \underline{74.69} & \underline{75.13} & 71.06 & \underline{73.24} & 70.59 & 75.29 & 78.00 & 99.44 & 22.20 & 86.42\\
& w / o Contrast & \underline{74.79} & 74.66 & 74.88 & \underline{72.52} & 72.89 & \textbf{72.91} & 15.26 & 15.71 & 99.86 & 82.82 & 96.83 \\
& w / o Cosine & 73.11 & 72.88 & 73.24 & 68.76 & 70.30 & 68.71 & \textbf{12.80} & \underline{13.14} & \underline{99.88} & \underline{83.14} & \textbf{97.29} \\
& w / o Binary & 73.93 & 73.75 & 73.92 & 71.07 & 72.31 & 70.98 & 13.92 & 14.29 & 99.86 & 78.06 & 96.83 \\
    & Full & \textbf{74.87} & \textbf{74.77} & \textbf{75.24} & \textbf{72.82} & \textbf{74.28} & \underline{72.42} & \underline{12.82} & \textbf{13.14} & \textbf{99.88} & \textbf{84.03} & \underline{97.19} \\
    \bottomrule 
\end{tabular}
}
\end{table*} 

\section{Results and Discussion}
\subsection{Main Results}
The main experimental results of ID classification and OOD detection are presented in Table~\ref{main_results}. Within each evaluation metric, the best performance is highlighted in bold and the second-best result is underscored. MIntOOD achieves superior performance across all three datasets, especially showing significant improvements in OOD detection performance, which demonstrates the efficacy and robustness of the learned multimodal representations in these tasks.

For ID classification, compared to the best-performing baselines, our method consistently outperforms them by approximately 1\% to 2\% on all metrics except recall on the MIntRec dataset. It also achieves improvements of more than 0.5\% on half of the evaluation metrics on the MELD-DA dataset and over 5\% on the precision metric. While SPECTRA stands out in recall on MELD-DA, our method outperforms it on the remaining five metrics with a significant margin (approximately 1\% to 9\%). On the IEMOCAP-DA dataset, our method secures state-of-the-art performance across five metrics, with only the precision metric being led by SDIF. It is evident that our method shows notably greater improvements on the MIntRec dataset than on the other datasets. This is likely because the intent categories in the MIntRec dataset contain more fine-grained and nuanced semantics compared to the coarse-grained communicative intents, showcasing our method's advantage in learning discriminative multimodal representations for intent classification. Moreover, two variants of our method, MIntOOD (R) and MIntOOD (T), exhibit strong and comparable performance on many metrics across the three datasets relative to the baselines. However, they underperform compared to MIntOOD, showing a decrease of over 2\% on the MIntRec dataset, underscoring the importance of non-verbal modalities in capturing complex intent semantics.

\begin{figure}
	\centering
	\includegraphics[scale=.4]{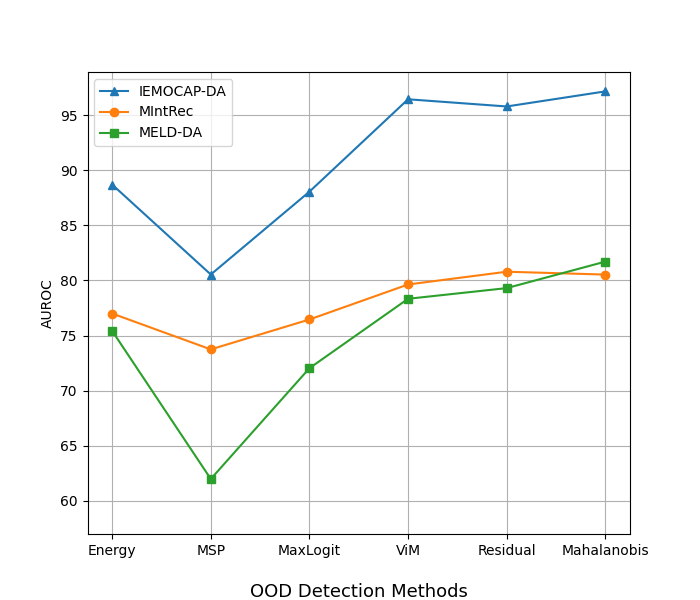}
 \caption{ A comparison between different OOD detection methods.}
	\label{ood_detection_methods}
\end{figure}

For OOD detection, our method demonstrates substantially greater improvements across all datasets. In particular, on the MIntRec dataset, our method shows improvements of approximately 3\% to 9\% across all  metrics compared with the best-performing baselines. On the MELD-DA dataset, although MMIM and SDIF are strong baselines with comparable performance on all metrics, our method still achieves more than 2\% improvement in AUPR-Out and AUROC. On the IEMOCAP-DA dataset, our method shows absolute advantages in DER, FPR95, AUPR-Out, and AUROC, with improvements of 44.85\%, 46.57\%, 61.49\%, and 8.36\%, respectively. The remaining metric, AUPR-In, reaches nearly 100\%, suggesting that there are fewer OOD samples in the IEMOCAP-DA test set compared to others, and that existing baselines tend to overfit the ID data while still assigning high scores to OOD data, leading to poorer overall performance. MIntOOD (R) achieves outstanding results across most metrics on all datasets, clearly surpassing the baselines. However, our full method maintains a decisive advantage, outperforming MIntOOD (R) by approximately 2\% to 8\%, 1\% to 3\%, and 0.5\% to 1\% on the respective datasets. These findings underscore the effectiveness of our OOD data generation strategy, which enhances the model's OOD detection capability even beyond that attained with real-world OOD samples. Compared to MIntOOD (T), our method shows substantial improvements in almost all metrics across the three datasets (with improvements of about 1\% to 3\%, 2\% to 4\%, and 2\% to 40\%, respectively), demonstrating that non-verbal modalities yield greater gains in detecting OOD samples.

 \begin{figure*}
  \centering
    \includegraphics[width=1\linewidth]{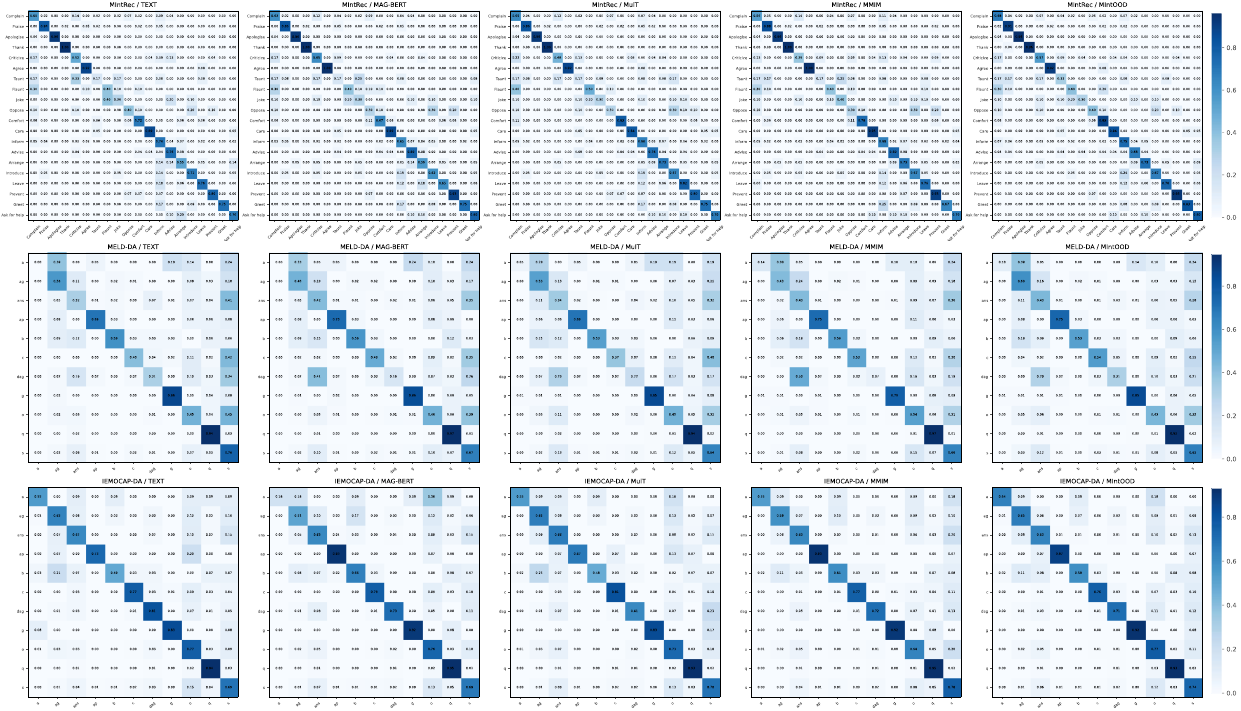}
  \caption{Confusion matrices for ID classes across the three datasets.}
  \label{fig:confusion}
\end{figure*}
\subsection{Ablation Studies}
\label{ablation_studies}

To validate the effectiveness of each component in MIntOOD, we conduct ablation studies, with results shown in Table~\ref{ablation_results}. First, we evaluate the weighted feature fusion network by comparing it to two standard multimodal fusion methods, namely Fusion (Add) and Fusion (Concat), as well as three fusion mechanisms from baselines: Fusion (MulT), Fusion (MAG-BERT), and Fusion (SDIF). Fusion (Add) sums the features of all three modalities, while Fusion (Concat) concatenates the features and projects them into a unified dimension $D_{\mathrm{H}}$ via a linear layer. The other three fusion methods use the fusion networks from their respective baselines. Our method consistently outperforms the two standard fusion methods by approximately 1\% to 5\% across most metrics for both ID classification and OOD detection on the MIntRec and MELD-DA datasets. It also shows a significant improvement of up to 6.22\% over the best fusion methods, particularly on the FPR95 metric of the MIntRec dataset. On the IEMOCAP-DA dataset, while the improvement in some ID classification metrics is marginal, our method achieves substantial gains of 9.1\%, 9.43\%, and 45.28\% on DER, FPR95, and AUPR-Out for OOD detection, respectively. These results demonstrate that the learned modality weights enhance multimodal fusion, leading to better representations than other strategies.

Second, we evaluate the impact of removing the contrastive loss (w/o Contrast). With the exception of ID classification on the IEMOCAP-DA dataset, removing the contrastive loss results in a performance drop of more than 1\% across most metrics for both tasks on all three datasets. This indicates that fine-grained learning of ID and OOD data interactions fosters more discriminative representation learning. However, the observed decrease is less significant compared to other ablation studies, suggesting that the contrastive loss plays an auxiliary role in enhancing multimodal representation learning.

Third, we replace the cosine classifier with a linear classifier (w/o Cosine), which leads to significant decreases of approximately 2\% to 6\% in ID classification performance across the three datasets, and reductions of 2\% to 8\% on the MIntRec and MELD-DA datasets. This reveals that the cosine classifier is crucial for effectively discriminating features and capturing confidence information in multimodal representations. Notably, in OOD detection on IEMOCAP-DA, the performance is comparable or even improves. . This is likely because the OOD data in IEMOCAP-DA contains distinct features (e.g., the absence of the text modality) that can be recognized by a simple linear layer without normalization.

Finally, we assess the impact of removing the binary training process (w/o Binary). This modification leads to a decrease of over 1\% in ID classification scores across all three datasets, with particularly notable reductions of 4\% to 6\% in F1-score, precision, and recall on the MELD-DA dataset. This component also plays a significant role in OOD detection, causing decreases of approximately 1\% to 7\% and 1\% to 4\% in all OOD metrics on the MELD-DA and MIntRec datasets, respectively. These significant reductions underscore the importance of distinguishing elementary binary information and leveraging it to guide representation learning.

\subsection{Effect of OOD Detection Methods}
\label{OOD_comparison}

To investigate the effect of different OOD detection methods, we compare our method, Mahalanobis (as introduced in Section~\ref{mahalanobis}), with five other state-of-the-art unsupervised OOD detection methods: 
\begin{itemize}
    \item \textbf{Energy}: This method applies the energy function~\cite{liu2020energy} to the output logits.
    \item \textbf{Residual}: As suggested in~\cite{zaeemzadeh2021out,wang2022vim}, we first offset the feature space to the origin and calculate the principal subspace using the largest eigenvalues (equal to the number of ID classes) of the covariance matrix. We then measure the deviation of the features extracted from the testing set from the principal subspace by computing the length of the features in the orthogonal complement of the principal subspace.
    \item \textbf{MSP}: This method calculates the maximum $\operatorname{Softmax}$ probabilities~\cite{hendrycks17baseline} by applying the $\operatorname{Softmax}$ function to the neural network output logits.
    \item \textbf{MaxLogit}: This approach involves direct maximization over the neural network output logits.
    \item \textbf{ViM}: Virtual-logit matching~\cite{wang2022vim} adjusts the residual score to generate a virtual logit using a scaling parameter, deriving from the ratio of the mean of the maximum training logits to the mean of the training residual scores. The ViM score is then computed by replacing the target logit of the original testing logits with the virtual logit. Before calculating the Residual and ViM scores, L2-normalization is applied to both the classifier weights and features, given our method's use of a cosine classifier.
\end{itemize}
\begin{figure*}
  \centering
    \includegraphics[width=1.01\linewidth]{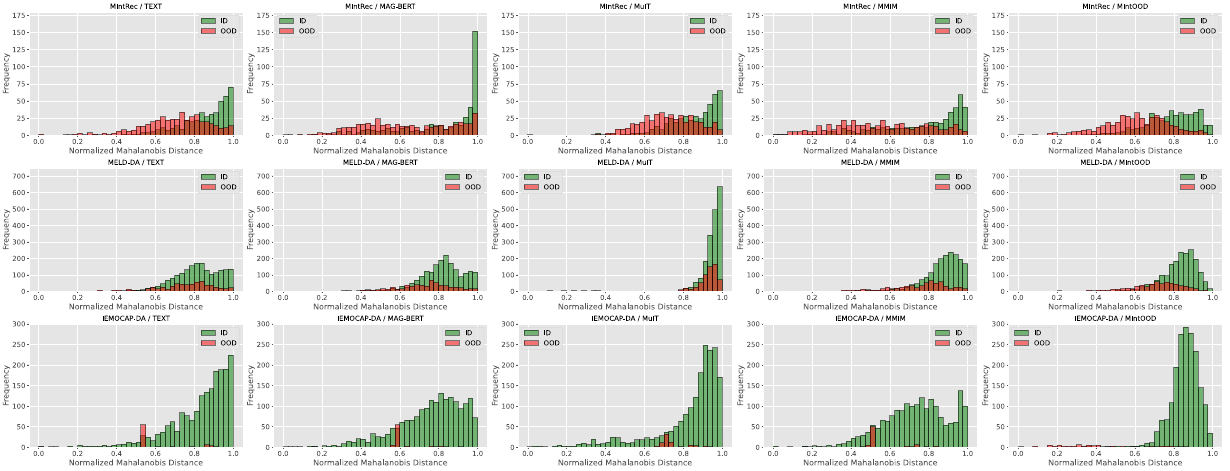}
    \vspace{-4mm} 
  \caption{Distribution of OOD detection scores for ID and OOD data in the testing sets of the three datasets.}
  \label{fig:confidence}
\end{figure*}

We use AUROC as the evaluation metric and present the results in Figure~\ref{ood_detection_methods}. Mahalanobis achieves the best performance on the IEMOCAP-DA and MELD-DA datasets and the second-best performance on the MIntRec dataset (slightly lower than Residual), demonstrating the effectiveness of using the Mahalanobis distance in the feature space to generate discriminative scores for OOD detection. It is observed that Energy, MSP, and MaxLogit consistently perform worse than the others. This is because they rely on the output logits from the cosine classifier, which tend to overfit to ID classes rather than capture the discrepancy between ID and OOD, thereby limiting their generalization ability on unseen OOD data. In contrast, ViM, Residual, and Mahalanobis utilize more original features prior to the cosine classifier, incorporating broader and more fundamental characteristics that might be lost or ignored after classification. Although Residual and ViM also focus on OOD detection in the feature space, they exhibit poorer performance than Mahalanobis. This is likely because Residual considers only the principal subspace, neglecting other dimensions that might also contribute to yielding discriminative scores. Additionally, ViM incorporates information from the output logit space, which may negatively impact the purity of the feature-space information.

\subsection{Analysis of ID Classification Performance}
To investigate the fine-grained performance of ID classification, we compare MIntOOD with four baselines using confusion matrices, as shown in Figure~\ref{fig:confusion}. Each value on the main diagonal denotes the accuracy for the respective class.

On the MIntRec dataset, our method achieves the best performance in 12 out of 20 classes, significantly outperforming the baselines. In contrast, the baselines TEXT, MulT, MAG-BERT, and MMIM achieve the best performance in only 4, 5, 6, and 7 classes, respectively. Specifically, our method shows improvements of 1\%, 2\%, 3\%, 6\%, 8\%, and 16\% over the best-performing baselines in the \textit{Inform}, \textit{Praise}, \textit{Complain}, \textit{Comfort}, \textit{Greet}, and \textit{Taunt} classes, respectively. It also leads in the \textit{Apologise}, \textit{Oppose}, \textit{Arrange}, \textit{Ask for help}, and \textit{Prevent} classes. These results demonstrate that our method effectively captures the nuanced features of ID classes, excelling in recognizing both straightforward intents (e.g., \textit{Greet}, \textit{Praise}, \textit{Apologise}) and those with complex semantics (e.g., \textit{Complain}, \textit{Taunt}, \textit{Ask for help}). For the remaining classes (e.g., \textit{Thank}, \textit{Care}, \textit{Flaunt}, \textit{Joke}, and \textit{Leave}), our method performs comparably to the baselines. The limited differentiation in classes such as \textit{Thank} and \textit{Care} is mainly because these classes rely heavily on textual cues, which gain little benefit from non-verbal modalities. Moreover, classes with intricate semantics, such as \textit{Flaunt}, \textit{Joke}, and \textit{Leave}, pose challenges in capturing high-level video or audio cues (e.g., expressions, body language, tones), which are crucial for intent understanding.

On the MELD-DA and IEMOCAP-DA datasets, our method achieves the best performance in 5 out of 11 ID classes and 4 out of 11 ID classes, respectively, maintaining its lead in the majority of categories. Specifically, on the MELD-DA dataset, MIntOOD excels in the \textit{Agreement} (ag), \textit{Answer} (ans), \textit{Apology} (ap), and \textit{Command} (c) classes, with improvements ranging from 1\% to 4\% over the best-performing baselines. Additionally, it exhibits comparable performance to the top baselines in the \textit{Disagreement} (dag), \textit{Greeting} (g), and \textit{Question} (q) classes, with differences within 2\%. On the IEMOCAP-DA dataset, MIntOOD leads in the \textit{Acknowledge} (a), \textit{Agreement} (ag), \textit{Greeting} (g), and \textit{Statement-Opinion} (o) classes, while also showing competitive performance in the \textit{Answer} (ans), \textit{Question} (q), and \textit{Statement-Non-Opinion} (s) classes. Notably, it significantly enhances the \textit{Acknowledge} class by 9\%, addressing the challenges posed by the limited textual information in responses that predominantly feature short forms. In such cases, our method leverages non-verbal modalities to boost intent recognition. Overall, these results indicate that our proposed method excels in distinguishing ambiguous and coarse-grained communicative intents by leveraging multimodal information. While it performs well in categories with clear linguistic features (e.g., \textit{Apology}, \textit{Greeting}, \textit{Question}), challenges remain in differentiating abstract dialogue acts, especially in classes such as \textit{Acknowledge}, \textit{Backchannel}, and \textit{Answer}, where performance often falls below 60\% due to similar semantics.

\subsection{Analysis of OOD Detection Performance}

To illustrate the model's capability to detect OOD data, we compute the confidence scores for each testing sample by normalizing the Mahalanobis distance scores $x_{\textrm{Ma}}$ within the range [0, 1] using min-max scaling: $x'_{\textrm{Ma}}=\frac{x_{\textrm{Ma}} - \min(x_{\textrm{Ma}})}{\max(x_{\textrm{Ma}}) - \min(x_{\textrm{Ma}})}$. We then plot the distribution of $x'_{\textrm{Ma}}$ for both ID and OOD data across the three datasets, as shown in Figure~\ref{fig:confidence}.

On the MIntRec dataset, it is observed that MIntOOD confines the confidence scores of OOD samples predominantly within the range [0.5, 0.7], maintaining a clear separation from the majority of ID sample scores, which lie in the range [0.7, 0.9]. This clear delineation indicates a distinct boundary between ID and OOD samples. In contrast, the baselines exhibit more overlap between ID and OOD samples. For instance, the confidence score ranges for overlapping ID and OOD samples are [0.7, 1.0] for TEXT, [0.75, 1.0] for MAG-BERT, [0.7, 0.9] for MulT, and [0.5, 0.9] for MMIM. Thus, MIntOOD shows superior generalization capabilities on unseen OOD data with more discriminative representations.

On the MELD-DA dataset, all baselines struggle to effectively distinguish between ID and OOD samples. Specifically, the differences between the peaks of the ID and OOD distributions are approximately 0, 0.075, 0.025, and 0.1 for TEXT, MAG-BERT, MulT, and MMIM, respectively. In contrast, MIntOOD shows a much larger separation of 0.175, clearly indicating fewer overlaps between OOD and ID samples compared to the baselines. However, some overlap remains in all methods, partly because the \textit{Others} class used as OOD data exhibits linguistic characteristics similar to those of ID data, making it difficult to differentiate due to ambiguous semantics.

On the IEMOCAP-DA dataset, MIntOOD exhibits exceptional performance by nearly completely separating ID from OOD data, with minimal overlap. In comparison, other methods show significant overlap, with over 50\% of OOD data overlapping with ID data. Despite the sparse presence of OOD data in the IEMOCAP-DA dataset, our method avoids overfitting to ID data and demonstrates remarkable robustness and discriminative capability, with most OOD scores below 0.5 and the majority of ID scores above 0.6. These results underscore the effectiveness of MIntOOD, which leverages specifically designed multimodal fusion and discriminative representation learning techniques to achieve strong generalization and reliability in detecting OOD data.

\section{Conclusions}
This paper proposes MIntOOD, a novel method designed to address the critical challenge of multimodal intent understanding by both accurately recognizing known intents and effectively detecting unseen OOD data. MIntOOD comprises two main modules. First, it introduces a simple yet effective multimodal fusion network that learns modality weights through dedicated neural networks. These learned weights are combined with the original encoded modality-specific features to produce robust multimodal representations. This strategy outperforms tensor operation-based or attention-based fusion methods, especially in OOD detection. Second, we develop discriminative representations for both tasks from three perspectives. After generating pseudo-OOD data from ID data, we initially learn coarse-grained features by distinguishing between binary ID and OOD classes. Subsequently, we eliminate the interference of vector magnitudes and use angular deviations from the cosine classifier to guide the differentiation of ID classes. Additionally, instance-level similarity relations are employed to further enhance discriminative representation learning. Each strategy plays a crucial role, and their removal results in substantial performance degradation, underscoring their effectiveness.

We establish baselines on three benchmark multimodal intent datasets, and an OOD benchmark is specifically built for the MIntRec dataset due to its previous absence. Extensive experiments demonstrate that our proposed method consistently achieves the best performance on ID classification across the three datasets, and significantly outperforms the baselines in OOD detection, with increases of up to 3\%$\sim$62\% on the AUPR-Out metric. Further analysis on ID classification and OOD detection provides additional evidence of the robustness and effectiveness of our method. We believe this work represents a significant advancement and serves as a pioneering success in this field.

\bibliography{mintood}
\bibliographystyle{IEEEtran}

\begin{IEEEbiography}  [{\includegraphics[width=1in,height=1.1in,clip,keepaspectratio]{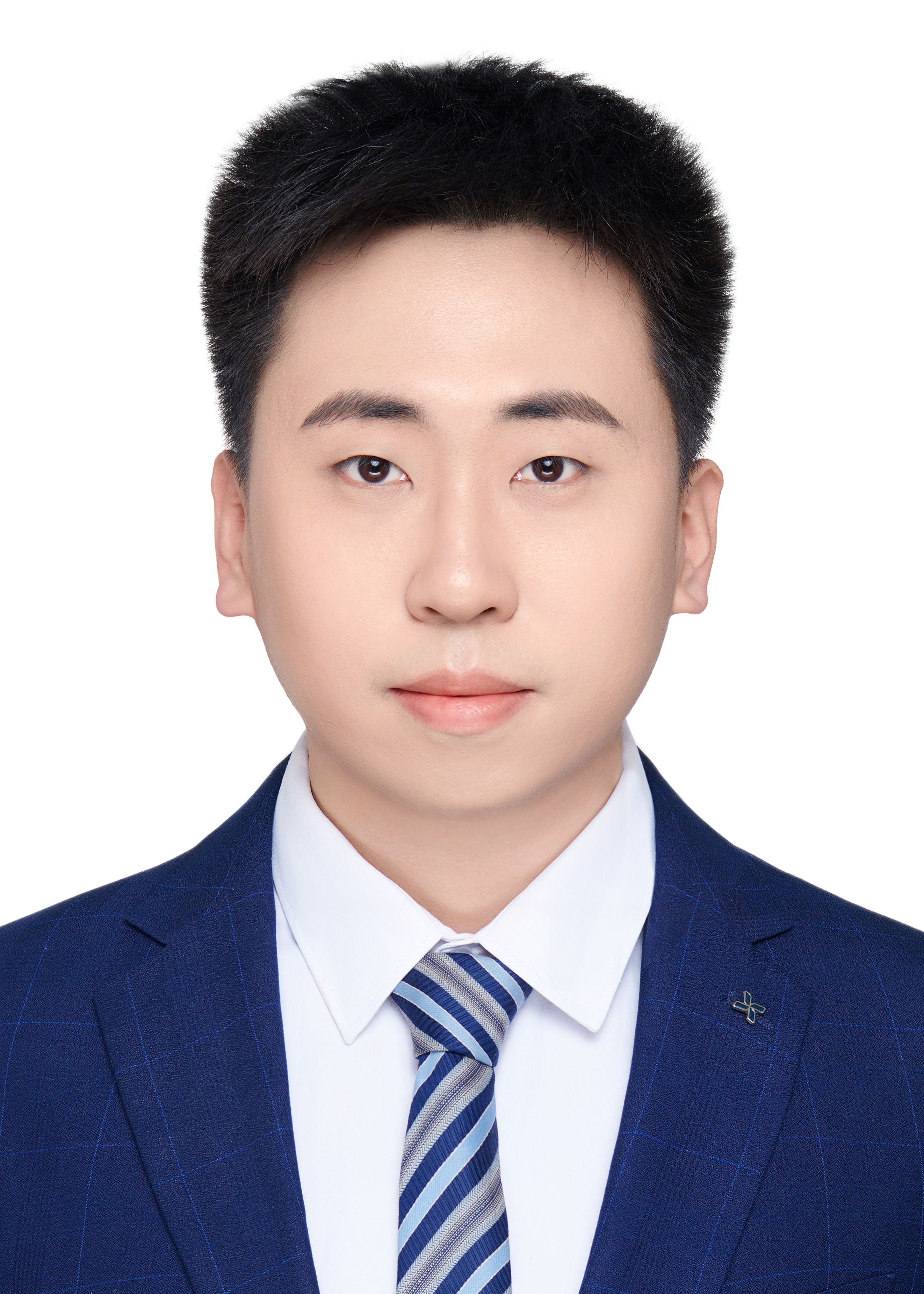}}]{Hanlei Zhang} received the B.S. degree from the Department of Computer Science and Technology, Beijing Jiaotong University, Beijing, China, in 2020. He is currently working toward the Ph.D. degree with the Department of Computer Science and Technology, Tsinghua University, Beijing, China. He has authored or coauthored 10 peer-reviewed papers, of which 8 are first-author papers in top-tier international conferences and journals, including ICLR, ACL, AAAI, ACM MM, IEEE Transactions on Knowledge and Data Engineering, and IEEE/ACM Transactions on Audio, Speech and Language Processing. His research interests include intent analysis, open world classification, clustering, multimodal machine learning, and large language models.
\end{IEEEbiography}

\begin{IEEEbiography} 	[{\includegraphics[width=1in,height=1.1in,clip,keepaspectratio]{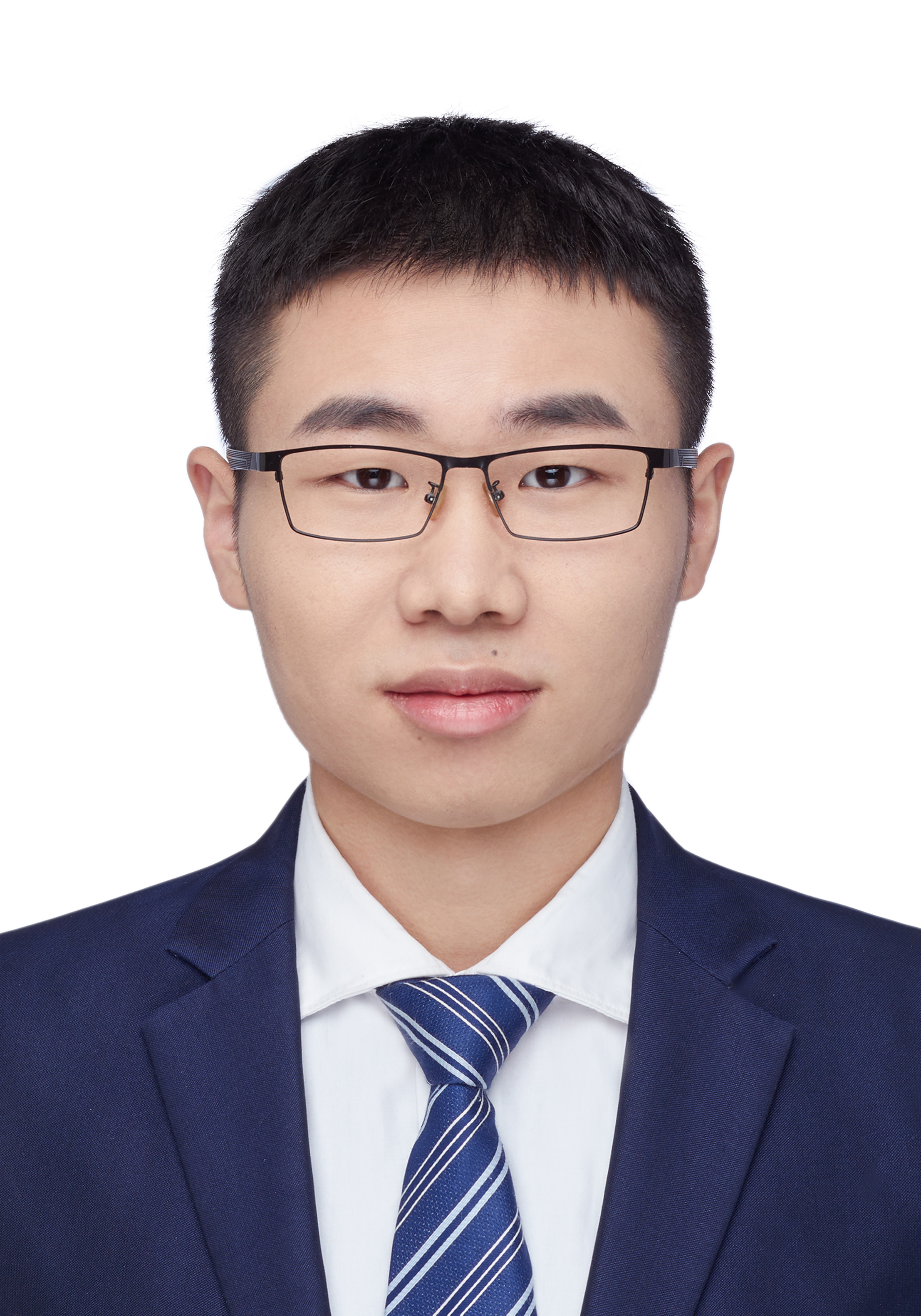}}]{Qianrui Zhou} received the B.S. degree in 2022 from the Department of Computer Science and Technology, Tsinghua University, Beijing, China, where he is currently working toward the Ph.D. degree with the Department of Computer Science and Technology. He has authored or coauthored five peer-reviewed papers in top-tier international venues, including ACL, ACM MM, AAAI, ICLR, and IEEE/ACM Transactions on Audio, Speech and Language Processing. His research interests include intent analysis, multimodal language understanding, and large language models.
\end{IEEEbiography}

\begin{IEEEbiography} 	[{\includegraphics[width=1in,height=1.1in,clip,keepaspectratio]{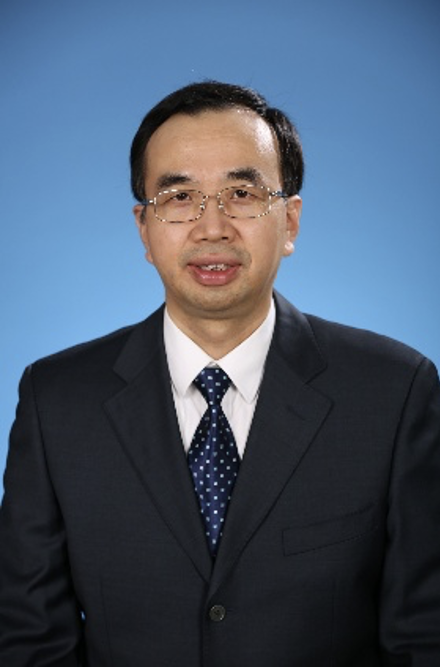}}]{Hua Xu} received his B.S. degree from Xi'an Jiao Tong University in 1998 and his M.S. and Ph.D degrees from Tsinghua University in 2000 and 2003 respectively. Dr. Xu is currently a Tenured Associate Professor in the Department of Computer Science at Tsinghua University. His research focuses on intelligent optimization and human-machine natural interaction in artificial intelligence. Dr. Xu has authored over 130 peer-reviewed papers, including 30 top international conference papers as the first or corresponding author and 45 high-impact SCI journal papers, accumulating 2000 SCI citations and 8900 Google Scholar citations. Additionally, Dr. Xu has authored 4 textbooks and 10 academic monographs. He holds 36 granted patents for inventions and 26 granted software copyrights. His contributions have been cited in top journals such as ACM Comput. Surv., with citations from 6 domestic and international academicians and over 30 IEEE Fellows. Dr. Xu has been invited to serve as the editor-in-chief of Elsevier's international journals Intell. Syst. Appl. and associate editor of Expert Syst. with Appl.. He has previously been honored with a National Science and Technology Progress Award, Beijing Science and Technology Award, two Industry Association Science and Technology Awards, and three second and third-class awards at the provincial and ministerial levels.
\end{IEEEbiography}

\begin{IEEEbiography} 	[{\includegraphics[width=1in,height=1.1in,clip,keepaspectratio]{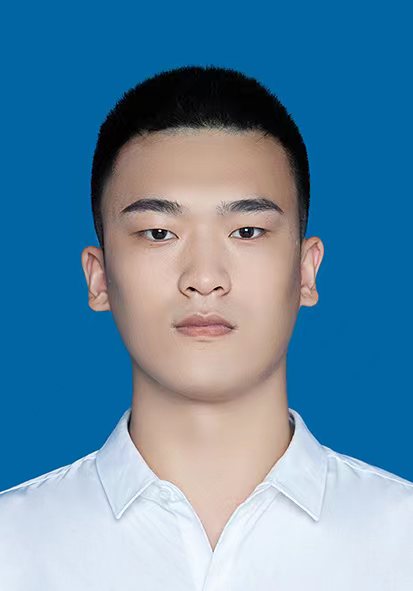}}]{Jianhua Su} is currently working toward the M.S. degree with the School of Information Science and Engineering. His research interests include natural language processing, multimodal fusion, and open intention detection.
\end{IEEEbiography}

\begin{IEEEbiography} 	[{\includegraphics[width=1in,height=1.1in,clip,keepaspectratio]{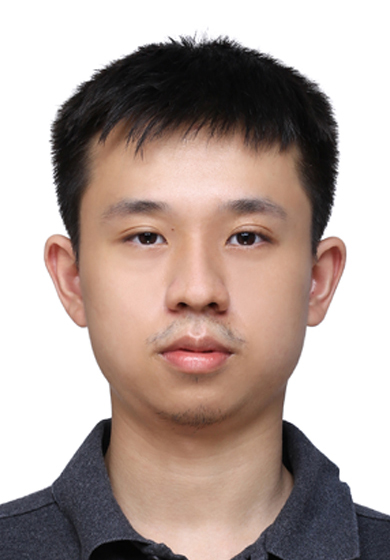}}]{Roberto Evans} received his B.S. degree from the Department of Computer Science and Technology, Beijing Institute of Technology, Beijing, China, in 2021 and his M.Sc. degree from the Department of Computer Science and Technology, Tsinghua University, Beijing, China, in 2024. His research interests include intent analysis, natural language processing, and multimodal learning. 
\end{IEEEbiography}

\begin{IEEEbiography} 	[{\includegraphics[width=1in,height=1.1in,clip,keepaspectratio]{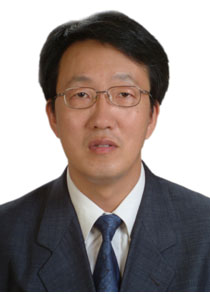}}]{Kai Gao} received the Ph.D. degree from Shanghai Jiao Tong University, Shanghai, China. He is a professor in the School of Information Science and Engineering, Hebei University of Science and Technology. He has authored or coauthored over 60 academic papers in international conferences and journals. His research interests include natural language processing, knowledge discovery, and multimodal intelligent information processing. 
\end{IEEEbiography}
\end{document}

%% file: Abstract.tex
\begin{abstract}
Multimodal intent understanding is a significant research area that requires effective leveraging of multiple modalities to analyze human language. Existing methods face two main challenges in this domain. Firstly, they have limitations in capturing the nuanced and high-level semantics underlying complex in-distribution (ID) multimodal intents. Secondly, they exhibit poor generalization when confronted with unseen out-of-distribution (OOD) data in real-world scenarios. To address these issues, we propose a novel method for both ID classification and OOD detection (MIntOOD). We first introduce a weighted feature fusion network that models multimodal representations. This network dynamically learns the importance of each modality, adapting to multimodal contexts. To develop discriminative representations for both tasks, we synthesize pseudo-OOD data from convex combinations of ID data and engage in multimodal representation learning from both coarse-grained and fine-grained perspectives. The coarse-grained perspective focuses on distinguishing between ID and OOD binary classes, while the fine-grained perspective not only enhances the discrimination between different ID classes but also captures instance-level interactions between ID and OOD samples, promoting proximity among similar instances and separation from dissimilar ones. We establish baselines for three multimodal intent datasets and build an OOD benchmark. Extensive experiments on these datasets demonstrate that our method significantly improves OOD detection performance with a 3\%$\sim$10\% increase in AUROC scores while achieving new state-of-the-art results in ID classification.\footnote{ Data and codes are available at~\url{https://github.com/thuiar/MIntOOD}.} 
\end{abstract}